\documentclass[default]{sn-jnl}% Default
\usepackage{graphicx}%
\usepackage{multirow}%
\usepackage{amsmath,amssymb,amsfonts}%
\usepackage{amsthm}%
\usepackage{mathrsfs}%
\usepackage[title]{appendix}%
\usepackage{xcolor}%
\usepackage{textcomp}%
\usepackage{manyfoot}%
\usepackage{booktabs}%
\usepackage{algorithm}%
\usepackage{algorithmicx}%
\usepackage{algpseudocode}%
\usepackage{listings}%
\usepackage{natbib}
\usepackage{subcaption}
\theoremstyle{thmstyleone}%
\newtheorem{theorem}{Theorem}%  meant for continuous numbers
\newtheorem{proposition}[theorem]{Proposition}% 
\usepackage{amsmath,epsfig,graphicx,setspace,url,float}
\theoremstyle{thmstyletwo}%
\usepackage[symbol]{footmisc}
\theoremstyle{thmstylethree}%
\newtheorem{lem}{Lemma}
\def\bomega{{\boldsymbol{\omega}}}

\def\b1{{\boldsymbol{1}}}
\def\c1{{\textcircled{a}}}

\def\bc{{\boldsymbol{c}}}
\def\bd{{\boldsymbol{d}}}

\def\bk{{\boldsymbol{k}}}

\def\bm{{\boldsymbol{m}}}

\def\bp{{\boldsymbol{p}}}

\def\bs{{\boldsymbol{s}}}

\def\bx{{\boldsymbol{x}}}
\def\by{{\mathbf{y}}}

\def\bI{{\mathbf{I}}}

\def\bX{{\boldsymbol{X}}}

\def\bZ{{\boldsymbol{Z}}}

\usepackage{cleveref}
\begin{document}

\title[Article Title]{Majorization-Minimization based Hybrid Localization Method  for High Precision Localization in Wireless Sensor Networks}


\author*[1]{\fnm{Kuntal} \sur{Panwar}}\email{Kuntal.Panwar@care.iitd.ac.in}

\author[1]{\fnm{Prabhu} \sur{Babu}}\email{Prabhu.Babu@care.iitd.ac.in}
\author[2]{\fnm{R} \sur{Jyothi}}\email{jyothi-rikhabchand@uiowa.edu}

\affil*[1]{\orgdiv{Centre for Applied Research in Electronics}, \orgname{Indian Institute of Technology, Delhi}, \orgaddress{\city{New Delhi}, \postcode{110016}, \country{India}}}
\affil[2]{\orgdiv{Department of Electrical and Computer Engineering}, \orgname{University of Iowai}, \orgaddress{\city{Iowa City}, \state{IA}, \postcode{5224}, \country{USA}}}

\abstract{This paper investigates the hybrid source localization problem using  the four radio measurements - time of arrival (TOA), time difference of arrival (TDOA), received signal strength (RSS), and angle of arrival (AOA). First, after invoking tractable approximations in the RSS and AOA models, the maximum likelihood estimation (MLE) problem for the hybrid TOA-TDOA-RSS-AOA data model is derived.  Then a weighted least-squares problem is formulated from the MLE, which is solved  using the principle of the majorization-minimization (MM), resulting in  an iterative algorithm with guaranteed convergence. The key feature of the proposed method is that it provides a unified framework where localization using any possible merger out of these four measurements can be implemented as per the requirement/application. Extensive numerical simulations are conducted  to study the performance of the proposed method.  The obtained results indicate that the hybrid localization model improves the localization accuracy compared to the heterogeneous measurements under different network scenarios, which also includes the presence of non-line of sight (NLOS) errors. }

\keywords{Hybrid source localization,  majorization-minimization (MM) algorithm, received signal strength (RSS), time of arrival (TOA), time-difference of arrival (TDOA),  angle of arrival (AOA).}
\footnotetext[2]{A preliminary version of the manuscript is accessible on arXiv with the following link: https://arxiv.org/abs/2205.03881v4}
\maketitle

\section{Introduction}\label{sec1}

Over the past few years, the problem of estimating the precise information  of source location has gained paramount interest  due to its utter requirement in various wireless communication applications  such as smart drones, e-healthcare, vehicular ad-hoc networks,  heterogeneous and internet-of-things networks, cellular networks, radio ranging, etc.
The high-precision localization of the source node system  enables various location and context-aware applications and helps to  assist/improve the data sampling/transfer, event scheduling, resource allocation and management, low-latency communication, etc. The ever-rising demand for low-latency communication, high-data-rate, ultra-reliable and dense architecture in future sensor networks has brought stringent requirements of localization accuracy in the order of one meter or even below. Although extensive research on localization methods has been carried out in the last decade or so, the investigation of high-precision localization solutions  is still largely open. % Importance of Accurate Localization in wireless communication
%Compared to the existing radio localization services such as cellular system based  uplink-time-difference of arrival (U-TDoA), global positioning system (GPS) or WiFi fingerprinting  based solutions yields localization accuracy in the order of few of meters. However, the localization accuracy of 5G networks is expected to be in the order of one meter or even below.  

Most commonly adopted localization solutions involve the extraction of range and angle information of the target from radio signals using time-of-arrival (TOA) \cite{toa_b}, time-difference-of-arrival (TDOA) \cite{tdoa_b,last1,last2}, received-signal-strength (RSS) \cite{rss_b} and angle-of-arrival (AOA) measurements \cite{aoa_b}. 
Recently, the hybrid localization methods involving effective and combined utilization of these heterogeneous measurements have gained
significant attention due to their improved localization accuracy compared with the standalone localization methods, i.e., methods utilizing one type of measurement. These hybrid methods can be categorized as  hybrid range-based methods and hybrid range-angle-based methods which involve the combination of range (TOA, TDOA, RSS) information and range and angle (AOA) information, respectively. The hybrid localization methods exploit the limited advantages of the heterogeneous measurement and can provide high-precision localization accuracy at best.
% Common ways to gain hyper accurate localization
%However, each of these exhibit limited advantages in terms of localization accuracy. For instance, RSS and AOA based localization solutions are generally preferred for short-range communication, while, TOA/TDOA measurements are preferred over long-range. AOA based localization solution exploit the merit of minimized-anchor connectivity as compared to RSS and TOA/TDOA. 

% Importance of sensor placement
\subsection{Related Work}
Owing to their improved localization accuracy and reliability, the usage of hybrid measurements for source localization  has been a research area of significant interest in the past few years. Many target localization algorithms were developed using different combinations of measurements, e.g., RSS-TOA/TDOA \cite{71,78,49,84,75,42,68,69,rss-tdoa,70,mkatwe,mine,last3},  TOA-AOA \cite{89,50,100,exsp1}, TDOA-AOA \cite{85,95,97,119}, RSS-AOA \cite{101,105,108,116,115,58,112,117,110,111,141,114,109}.

The hybrid range-based localization methods that combine the TOA/TDOA and RSS measurements are mostly utilized in sensor network applications, which will be briefly reviewed in the following.  The  Cramer-Rao lower bound (CRLB) corresponding to fused TOA-RSS measurements was derived in \cite{71}.  In \cite{78}, a technique based on the Newton-Raphson method was proposed to solve the hybrid  TOA-RSS maximum likelihood  estimation (MLE) problem. The authors of \cite{49}  proposed a weighted least-squares (WLS) method obtained via an ad-hoc relaxation of the MLE  problem for the hybrid TOA-RSS measurement model. In \cite{84}, source localization for an indoor environment using the hybrid TOA-RSS model was proposed  where the line of sight (LOS)/non-line of sight (NLOS) paths were pre-identified.  The authors of \cite{75} proposed a two-step WLS method for localization in a cooperative  network  where TOA and RSS measurements were combined, and it was shown to  outperform localization estimates obtained via standalone measurements. A two-stage algorithm for localization with unknown channel parameters was proposed in \cite{42} for the hybrid TOA-RSS model; with the first stage performing the calibration of model parameters through known positions of the anchor nodes, followed by an iterative approach in the second stage to arrive at the source location.

The hybrid TOA-RSS measurements are also found to be effective in the mitigation of NLOS errors \cite{68,69,rss-tdoa,70,mkatwe,mine,last3}. A generalized trust region sub-problem (GTRS) based on the MLE approach, with a balancing parameter included for modeling the NLOS errors, was developed for the hybrid TOA-RSS model in \cite{68}. A similar but robust approach, called RGTRS \cite{69}, was proposed based on the worst-case approximation technique for the TOA-RSS model. Similarly, the GTRS framework for RSS-TDOA measurements was developed in \cite{rss-tdoa} using the bisection approach. The methods in \cite{68,69,rss-tdoa} proved the efficiency of localization through hybrid measurements over individual as well as double measurements.  In \cite{70}, the network topology of the hybrid TOA-RSS model was explored to improve the localization accuracy by first estimating the azimuth angle through available range observations and then amalgamating the range-angle observations and a closed-form solution of the location was obtained.  The authors of \cite{mkatwe} designed a non-linear weighted least-squares (NLWLS) objective from the MLE problem of the hybrid TOA-RSS model and solved it as a semi-definite programming (SDP) problem. Although the mean square error (MSE) of the SDP-based approach was close to the CRLB, the method was found to be computationally demanding. This issue of increased computational complexity was addressed in \cite{mine}, where a minimizer of the NLWLS objective was obtained using a simple iterative algorithm.

Next, we will review the range-angle-based hybrid methods. The range-angle-based  techniques are advantageous in low anchor density networks. Among these, the hybrid TOA-AOA and TDOA-AOA measurements based localization algorithms are generally adopted for cellular networks and radar systems where there is the availability of ultra wide-band (UWB) receivers (for precise TOA measurements) and smart antennas (for precise AOA measurements). Whereas, the hybrid RSS-AOA based localization methods are commonly used in narrow-band low cost and low-range applications like WSNs and indoor communication applications. First, we will discuss the existing literature on  TOA-AOA and TDOA-AOA measurements based localization methods. The CRLB for these hybrid models was derived in \cite{89}.	In \cite{50}, localization in a cooperative network was obtained using the WLS method with weights chosen based on  the noise variances of the measurements corresponding to the TOA-AOA  model. A cooperative distributed localization algorithm was presented for the TOA-AOA model in \cite{100}, where the proposed message-passing hybrid localization (MPHL) algorithm was derived based on belief propagation and Markov Chain Monte Carlo sampling technique. Target location estimation methods for the TDOA-AOA hybrid model were derived through the WLS approach in \cite{85,95}; however, these methods suffered from significant bias. The large bias incurred in the aforementioned methods was reduced using the structured total least square (STLS) approach in \cite{97}, and the method was also found to be effective when the target was present outside the convex hull formed by the sensor locations. In \cite{119}, a target localization approach was proposed for TDOA-AOA measurements in polar coordinates for both near and far-field regions.

Finally, we will review the literature on localization methods  based on hybrid RSS-AOA measurements. The hybrid approaches  based on the RSS-AOA model are categorized on the basis of the availability of prior knowledge of the transmission power in the RSS measurements.  For known transmission power, localization using RSS-AOA measurements was mostly solved through the WLS method \cite{101,105,108,116}. For instance, in \cite{105}, a WLS based method employing an approximate estimate of a noise covariance matrix was proposed for localization when the prior information on noise statistics is unknown.  In \cite{115}, in the case of known transmission power, a least-squares (LS) based formulation  was proposed, and it was solved  as a second-order cone program (SOCP). The authors of \cite{58} dealt with the differential RSS and AOA measurement model and proposed a selective hybrid measurement weighted instrumentation variable (SHM-WIV) method.  Similar to  the case of known transmitted power, the models based on unknown transmission power were also generally solved using the WLS approach \cite{112,117,110,111,141,114,109}; in which the non-convex WLS  problem was usually relaxed to a convex problem and solved as either SDP \cite{117,110} or as SOCP \cite{111,141,114}.

\subsection{Motivation and Contribution}

Albeit the existing works distinctively show that the hybrid localization methods (employing two combinations such as TOA-RSS, RSS-AOA, TOA-AOA, etc)
provide better localization accuracy, the joint consideration of all four measurements and  overall analysis for high-precision localization is still an interesting and unexplored topic of research. To the best of our knowledge, a hybrid method combining all  four measurements (TOA, TDOA, RSS and AOA) has not yet been proposed. In this paper, a source localization algorithm is proposed for the hybrid TOA-TDOA-RSS-AOA measurement model. The major contributions of
this paper are detailed below:
\begin{itemize}
    \item The hybrid TOA-TDOA-RSS-AOA model is described, and the design objective is formulated using some general approximations.  
    
\item The proposed algorithm is developed using the principle of the Majorization-Minimization (MM) \cite{MMalgo}. The MM-based framework results in an iterative algorithm that  monotonically decreases the design objective with guaranteed  convergence to a stationary point of the design objective.
\item The framework presented is flexible to handle the fusion of any range and angle measurements. This provides an additional advantage of including or excluding any type of measurement based on the application in hand.
\item We present comprehensive numerical simulations using all the possible combinations of the proposed method and show its efficiency over a conventionally used approach. 
\end{itemize}
\subsection{Paper Notations and Structure}
\textit{Notations:} Throughout the paper, the scalars, vectors, and matrices are represented by regular (e.g., $x,\text{ }X$), bold lowercase (e.g., $\bx$) and bold uppercase (e.g., $\bX$), respectively. The $i^{\text{th}}$ element of vector $\mathbf{x}$ and the $(i,j)^{\text{th}}$
element of the matrix $\mathbf{X}$ are denoted as
$x_{i}$ and $X_{i,j}$, respectively. $||\bx||$ indicates the $\ell_2$ norm of the vector $\bx$. The notations \(\ln\) and \(\log\) denote the natural logarithm and logarithm with base 10, respectively. The notations $\arctan(x)$ and $\arccos(x)$ denote the arctangent and arccosine value of $x$, respectively. $ x^{(t)}$ denotes the value of parameter $x$ at the $t^{\text{th}}$ iteration. $\mathbb{N}$ denotes the set of
natural numbers and $\mathbb{E}(X)$ represents the expectation of 
variable $X$. The first  and second-order derivatives with respect to $x$ are given by $\frac{\partial(\cdot)}{\partial x}$ and $\frac{\partial^2(\cdot)}{\partial x^2}$, respectively. The signum (or sign) function is represented by $\text{sgn}(\cdot)$.   The probability density function of the random variable $\mathbf{x}$ is denoted by $p\left(\mathbf{x}\right)$. The transpose and inverse of a matrix are denoted as $\mathbf{X}^{T}$ and $\mathbf{X}^{-1}$, respectively.  $\mathbf{I}$ is used to represent an identity matrix of appropriate size. A diagonal matrix is denoted by diag$(\bx)$ where the diagonal elements are the corresponding elements of vector $\bx$. The operation $\odot$ denotes the element-wise multiplication of two vectors. $x\sim\mathcal{N}(\mu,\sigma^2)$ indicates that $x$ is Gaussian random variable with a mean of $\mu$ and variance of $\sigma^2$ and $\mathcal{U}\left(a,b\right)$ denotes uniform distribution taking
values between $a$ and $b$. 

\textit{Structure:} The rest of the paper is organized as follows. Section II describes the hybrid TOA-TDOA-RSS-AOA measurement model and problem formulation for source localization using the hybrid model. In Section III, we present the proposed approach based on the MM framework.  Simulation results  are presented  in Section IV, and finally, Section V concludes the paper.

\section{Measurement Model and Hybrid Localization Problem Formulation}
\begin{figure}[!t]
    \centering
    \includegraphics[scale=0.5]{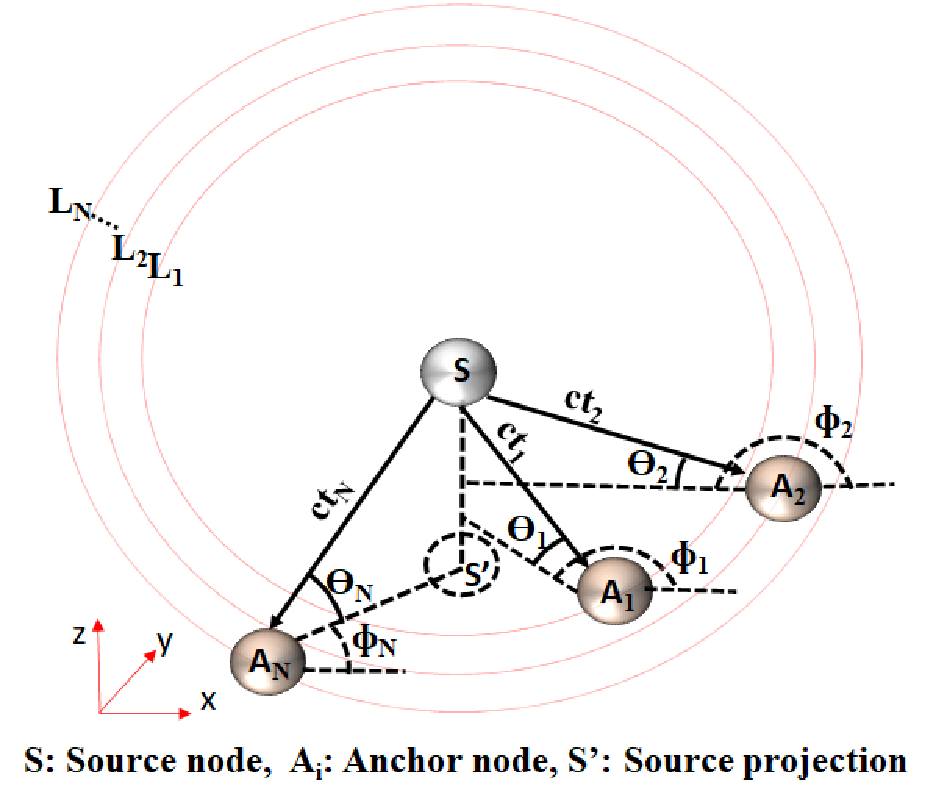}
    \caption{\text{3D WSN}: $ct_i$ - distance between source node and $i^{\text{th}}$ anchor node, $L_i$ - path loss at $i^{\text{th}}$ anchor node and $\phi _i$ and $\theta_i$ - azimuth and elevation angle between source node and $i^{\text{th}}$ anchor node $\forall i$}
    \label{1}
\end{figure}

Consider an $n$-dimensional ($n=$ 2 or 3) sensor  network that consists of $N$ anchors and an unknown source node. A representation of one such sensor network in  3D is shown in Figure \ref{1}. Let the true and known position of the $i^{\text{th}}$ anchor be defined as $\bm_i=[m_{x_i},m_{y_i},m_{z_i}]^T\in\mathbb{R}^3,i\in \mathbb{N}_n\triangleq\{1,\dots,N\}$, whereas the true but unknown position of source node be defined as  $\bs=[s_x,s_y,s_z]^T\in\mathbb{R}^3$.

Let us assume that the source node transmits a radio signal and each anchor extracts the angle/range or both information using radio  measurements on the received signal. With this the path loss $L_i$ of signal  measured at $i^{\text{th}}$ anchor can be related with range information as 
\begin{equation}\label{rss1}
    \begin{array}{ll}
   \text{RSS: }    L_{i}= L_{0} +10\gamma\log_{10}{\left\lvert\left\lvert\bs-{\bm}_i\right\rvert\right\rvert}+ {n}_{{\scriptscriptstyle\text{RSS}}_{i}},i\in\mathbb{N}_n,       \end{array}
\end{equation}
where $L_{0}$ is the path loss of the signal at reference distance $d_0$ which is assumed to be 1m from the source, $\gamma$ denotes the path loss exponent and
$n_{{\scriptscriptstyle\text{RSS}}_i}$ is the noise in RSS (or power\footnote{The path loss model is related to received signal power measurement such that $L_i=10 \log_{10}(\frac{P_t}{P_i})$, where $P_t$ and $P_i$ are, respectively, transmitted signal power and received power between two source and anchor nodes.}) measurement which represents log-normal shadowing and modeled   as zero-mean Gaussian random variable, i.e. $n_{{\scriptscriptstyle\text{RSS}}_i}\sim\mathcal{N}(0,\sigma_{{\scriptscriptstyle\text{RSS}}_i}^2)$. The RSS measurement model in \eqref{rss1} is non-linear due to the presence of $\log_{10}{\left\lvert\left\lvert\bs-{\bm}_i\right\rvert\right\rvert}$ term, and it is usually approximated (for tractability) as follows. The model equation in \eqref{rss1} can be re-arranged as
\begin{equation}\label{rss2}
 10^{\frac{L_o-L_i}{10\gamma}}\left\lvert\left\lvert\bs-{\bm}_i\right\rvert\right\rvert=10^{-\frac{n_{{\scriptscriptstyle\text{RSS}}_i}}{10\gamma}},\forall i   
 \end{equation}
For sufficiently low noise ($n_{{\scriptscriptstyle\text{RSS}}_i}\leq10\gamma$), the right side
expression of \eqref{rss2} can be approximated as $1-\frac{n_{{\scriptscriptstyle\text{RSS}}_i}}{\eta}$, where $\eta =\frac{10\gamma}{\log(10)}$, 
using first-order Taylor series expansion, thus \eqref{rss2} can be approximated as
\begin{equation}\label{rss3}
\eta-\eta\lambda_i\left\lvert\left\lvert\bs-{\bm}_i\right\rvert\right\rvert\approx n_{{\scriptscriptstyle\text{RSS}}_i},    \end{equation}
where $\lambda_i=10^{\frac{L_o-L_i}{10\gamma}}$. From hereafter, for the RSS measurements, we will work only with the approximated model in \eqref{rss3}. 

The TOA and TDOA\footnote{In the presence of TOA measurements, adding TDOA measurements is redundant as it does not add any additional new information regarding the source location. However, by working with the time differences, some of the unobserved errors (such as clock synchronization errors) in the TOA model equations are canceled out in the TDOA model. Only for the above-mentioned reason do we consider both TOA and TDOA models in our framework.} measurement at $i^{\text{th}}$ anchor can be expressed as
\begin{equation}\label{toa}
    \begin{array}{ll}
  \text{TOA: }   ct_i=\left\lvert\left\lvert\bs-{\bm}_i\right\rvert\right\rvert+n_{{\scriptscriptstyle\text{TOA}}_i},i\in\mathbb{N}_n,   
    \end{array}
\end{equation}
\begin{equation}\label{tdoa}
    \begin{array}{ll}
 \text{TDOA: }    c\left(t_{i}-t_{1}\right)=\left\lvert\left\lvert\bs-{\bm}_{i}\right\rvert\right\rvert-\left\lvert\left\lvert\bs-{\bm}_1\right\rvert\right\rvert+n_{{\scriptscriptstyle\text{TDOA}}_i},i=2,\dots,N, \end{array}
\end{equation}
where $c$ is the speed of the radio signal, $t_i$ is the propagation delays of the signal received at $i^{\text{th}}$ anchor. The first anchor is taken as a reference anchor for calculating TDOA measurements.   The quantities  $n_{{\scriptscriptstyle\text{TOA}}_i}\sim\mathcal{N}(0,\sigma_{{\scriptscriptstyle\text{TOA}}_i}^2)$ and $n_{{\scriptscriptstyle\text{TDOA}}_i}\sim\mathcal{N}(0,\sigma_{{\scriptscriptstyle\text{TDOA}}_i}^2)$ are the time measurement noises for TOA and TDOA, respectively which accounts for errors, if any and $\sigma_{{\scriptscriptstyle\text{TOA}}_i}^2$ is the noise variance of time measurement.    

Further, the 3D AOA measurements at $i^{\text{th}}$ anchor can be given as
\begin{align}
\phi_i=\arctan\left(\frac{s_y-m_{y_i}}{s_x-m_{x_i}}\right)+\bar{{n}}_{{\scriptscriptstyle\text{AOA}}_i},i\in\mathbb{N}_n,\label{aoa1}\\[-15pt]
\intertext{ AOA: \vskip-15pt}
\theta_i=\arccos\left(\frac{s_z-m_{z_i}}{\left\lvert\left\lvert\bs-{\bm}_i\right\rvert\right\rvert}\right)+\bar{\bar{{n}}}_{{\scriptscriptstyle\text{AOA}}_i},i\in\mathbb{N}_n,\label{aoa2}
\end{align}
where  $\phi_i$ and $\theta_i$ are the measured azimuth and elevation angle at the $i^{\text{th}}$ anchor and $\bar{n}_{{\scriptscriptstyle\text{AOA}}_i}$ and $\bar{\bar{n}}_{{\scriptscriptstyle\text{AOA}}_i}$ are the associated AOA measurement noises. Under low noise conditions, the AOA measurements in \eqref{aoa1} and \eqref{aoa2} can be written in a pseudo-linear form (this again done for tractability) as 
\begin{equation}
     \begin{array}{ll}
      &{\bc}_i^T\left(\bs-\bm_i\right)=\bar{n}_{{\scriptscriptstyle\text{AOA}}_i},\label{aoam1}
      \end{array}
      \end{equation}
      \begin{equation}
      \begin{array}{ll}
    &\bk^{T}(\bs -\bm_i)- \|\bs-\bm_{i}\|\cos \theta_{i}=\bar{\bar{n}}_{{\scriptscriptstyle\text{AOA}}_i}\label{aoam2},    
     \end{array}
 \end{equation}
where $\bk=[0,0,1]^T$ and $\bc_i=[-\sin\phi_i,\cos\phi_i,0]^T$. In \eqref{aoam1} and \eqref{aoam2}, we assume both $\bar{n}_{{\scriptscriptstyle\text{AOA}}_i}$ and $\bar{\bar{n}}_{{\scriptscriptstyle\text{AOA}}_i}$ to be Gaussian distributed with zero-mean and variance $\sigma_{{\scriptscriptstyle\text{AOA}}_i}^2$. 

 From \eqref{rss1}, \eqref{toa}, \eqref{tdoa}, \eqref{aoam1} and \eqref{aoam2}, the 
 the problem of least squares estimation of source node location can be given by
\begin{equation}\label{jml}
    \begin{array}{ll}
      \hat{\bs}=\arg\min\limits_{\bs}\sum\limits_{i=1}^{N}\left\lbrace \frac{f_{{\scriptscriptstyle\text{RSS}}_i}(\bs)}{\sigma_{{\scriptscriptstyle\text{RSS}}_i}^2}+\frac{f_{{\scriptscriptstyle\text{TOA}}_i}(\bs)}{\sigma_{{\scriptscriptstyle\text{TOA}}_i}^2}+\frac{f_{{\scriptscriptstyle\text{AOA}}_i}(\bs)}{\sigma_{{\scriptscriptstyle\text{AOA}}_i}^2}\right\rbrace+\sum\limits_{i=2}^{N}\left\lbrace \frac{f_{{\scriptscriptstyle\text{TDOA}}_i}(\bs)}{\sigma_{{\scriptscriptstyle\text{TDOA}}_i}^2}\right\rbrace,  
    \end{array}
\end{equation}
where
\begin{equation}\label{lik_rss}
    \begin{array}{ll}
     f_{{\scriptscriptstyle\text{RSS}}_i}(\bs)=\left(\eta-\eta\lambda_i\left\lvert\left\lvert\bs-{\bm}_i\right\rvert\right\rvert\right)^2,    
    \end{array}
\end{equation}
\begin{equation}\label{lik_toa}
    \begin{array}{ll}
     f_{{\scriptscriptstyle\text{TOA}}_i}(\bs)={\left(\tau_{i}\!-\!\left\lvert\left\lvert\bs-{\bm}_i\right\rvert\right\rvert\right)^2},    
    \end{array}
\end{equation}
\begin{equation}\label{lik_tdoa}    
    \begin{array}{ll}
     f_{{\scriptscriptstyle\text{TDOA}}_i}(\bs)={\left(\tau_{1i}\!-\! \left\lvert\left\lvert\bs-{\bm}_{i}\right\rvert\right\rvert+\left\lvert\left\lvert\bs-{\bm}_1\right\rvert\right\rvert\right)^2},
    \end{array}
\end{equation}
\begin{equation}\label{lik_aoa}
    \begin{array}{ll}
     f_{{\scriptscriptstyle\text{AOA}}_i}(\bs)=\left({\bc}_i^T\left(\bs-\bm_i\right)\right)^2 + \left(\bk^{T}(\bs -\bm_i)- \|\bs-\bm_{i}\|\cos \theta_{i}\right)^2 .    
    \end{array}
\end{equation}

\section{Proposed Algorithm}
In this section, we present the proposed localization algorithm that minimizes the MLE in \eqref{jml}. Before we proceed  to solve \eqref{jml}, we will present a weighing scheme to weigh different types of measurements which can  help to interplay with the  range-based characteristics of the different types of measurements. 
\subsection{Measurements Weighing}
Here, we define the range-based performance of all the considered methods through the following proposition. 
\begin{proposition}\label{wt_prop}
The radio measurements exhibit range-based characteristics. Like, the RSS and AOA  perform better at short ranges, and their localization accuracy degrades as range increases compared to TOA and TDOA measurements.
\end{proposition}
\begin{proof}
The CRLB provide minimum achievable variance on the parameter estimation. We here derive the CRLB expression for range/range-difference\footnote{Since we are providing range-dependent relation of measurements, we calculate the CRLB expression only in terms of ranges/range-differences and not the actual source location.} estimation  for all the measurements. The distance between the source and the $i^{\text{th}}$ anchor is given by $d_i=\|\bs-\bm_i\|$ and similarly $d_{f_i}=\|\bs-\bm_i\|-\|\bs-\bm_1\|$ denote the range difference between $i^{\text{th}}$ anchor and the reference sensor. Using \eqref{lik_rss}, \eqref{lik_toa} and \eqref{lik_tdoa}, the CRLB\footnote{In the CRLB derivation for RSS and AOA models, the functions $f_{{\scriptscriptstyle\text{RSS}}_i}$ and $f_{{\scriptscriptstyle\text{AOA}}_i}$ are derived from original data models without any approximations, as given in \cref{lik_rss,lik_toa,lik_tdoa}.} for range estimation can be given as \cite{49}
\begin{equation}\label{crb_rss}
    \begin{array}{ll}
     \text{CRLB}_{\scriptscriptstyle\text{RSS}}(d_i)=-\frac{1}{\frac{1}{2\sigma_{{\scriptscriptstyle\text{RSS}}_i}^2}\mathbb{E}\left[\frac{\partial^2 f_{{\scriptscriptstyle\text{RSS}}_i} }{\partial d_i^2}\right]}=\frac{\sigma_{{\scriptscriptstyle\text{RSS}}_i}^2d_{i}^2}{\eta^2},
     \end{array}
\end{equation}
\begin{equation}\label{crb_toa}
    \begin{array}{ll}
     \text{CRLB}_{{\scriptscriptstyle\text{TOA}}}(d_i)=-\frac{1}{\frac{1}{2\sigma_{{\scriptscriptstyle\text{TOA}}_i}^2}\mathbb{E}\left[\frac{\partial^2 f_{{\scriptscriptstyle\text{TOA}}_i} }{\partial d_i^2}\right]}=\sigma_{{\scriptscriptstyle\text{TOA}}_i}^2,   
    \end{array}
\end{equation}
\begin{equation}\label{crb_tdoa} 
    \begin{array}{ll}
     \text{CRLB}_{{\scriptscriptstyle\text{TDOA}}}(d_{f_i})=-\frac{1}{\frac{1}{2\sigma_{{\scriptscriptstyle\text{TDOA}}_i}^2}\mathbb{E}\left[\frac{\partial^2 f_{{\scriptscriptstyle\text{TDOA}}_i} }{\partial d_{f_i}^2}\right]}=\sigma_{{\scriptscriptstyle\text{TDOA}}_i}^2,   
    \end{array}
\end{equation}
respectively. In case of AOA, we have
\begin{equation}
    \begin{array}{ll}
       \text{CRLB}_{\scriptscriptstyle\text{AOA}}(d_i)=-\frac{1}{\frac{1}{2\sigma_{{\scriptscriptstyle\text{AOA}}_i}^2}\mathbb{E}\left[\frac{\partial^2 f_{{\scriptscriptstyle\text{AOA}}_i} }{\partial d_i^2}\right]}=\frac{\sigma_{{\scriptscriptstyle\text{AOA}}_i}^2}{\left(\frac{\partial f_{{\scriptscriptstyle\text{AOA}}_i} }{\partial d_i}\right)^2},  
    \end{array}
\end{equation}
where
\begin{equation}
    \begin{array}{ll}
     \frac{\partial f_{{\scriptscriptstyle\text{AOA}}_i} }{\partial d_i}=\left(\frac{\partial f_{{\scriptscriptstyle\text{AOA}}_i} }{\partial s_x}\middle/\frac{\partial d_i }{\partial s_x}\right)+\left(\frac{\partial f_{{\scriptscriptstyle\text{AOA}}_i} }{\partial s_y}\middle/\frac{\partial d_i }{\partial s_y}\right)+\left(\frac{\partial f_{{\scriptscriptstyle\text{AOA}}_i} }{\partial s_z}\middle/\frac{\partial d_i }{\partial s_z}\right).    
    \end{array}
\end{equation}
It can be obtained\footnote{Due to space brevity, the complete derivation is not included here.} that 
\begin{equation}
    \begin{array}{ll}
      \frac{\partial f_{{\scriptscriptstyle\text{AOA}}_i} }{\partial d_i}=-\frac{1}{d_i}&\left(\frac{2\cot2\phi_i}{\sin^2\theta_i}+{2\cot\theta_i-\tan\theta_i}\right),   
    \end{array}
\end{equation}
Therefore,
\begin{equation}\label{crb_aoa}
    \begin{array}{ll}
      \text{CRLB}_{{\scriptscriptstyle\text{AOA}}}(d_i)\propto \sigma_{{\scriptscriptstyle\text{AOA}}_i}^2d_i^2.   
    \end{array}
\end{equation}
It can be noted from the expressions in \eqref{crb_rss}, \eqref{crb_toa}, \eqref{crb_tdoa} and \eqref{crb_aoa} that the CRLBs of the TOA and TDOA are ``range independent" and the CRLBs of RSS and AOA are directly proportional to the range.
\end{proof}
  Thus, it is really important to choose the weights for the different types of measurements accordingly. For instance, one possible way to choose weights can be as follows:
 \begin{equation}\label{wts}
     \begin{array}{ll}
      &w_{{\scriptscriptstyle\text{RSS}}_i}=\dfrac{1}{\sigma_{{\scriptscriptstyle\text{RSS}}_i}^2}\left(1-\frac{e_{{\scriptscriptstyle\text{RSS}}_i}^2}{\sum\limits_{i=1}^{N} e_{{\scriptscriptstyle\text{RSS}}_i}^2}\right),\\
      &w_{{\scriptscriptstyle\text{TOA}}_i}=\dfrac{1}{\sigma_{{\scriptscriptstyle\text{TOA}}_i}^2}\left(1-\frac{e_{{\scriptscriptstyle\text{TOA}}_i}^2}{\sum\limits_{i=1}^{N} e_{{\scriptscriptstyle\text{TOA}}_i}^2}\right),\\
&w_{{\scriptscriptstyle\text{TDOA}}_i}=\dfrac{1}{\sigma_{{\scriptscriptstyle\text{TDOA}}_i}^2}\left(1-\frac{e_{{\scriptscriptstyle\text{TDOA}}_i}^2}{\sum\limits_{i=2}^{N} e_{{\scriptscriptstyle\text{TDOA}}_i}^2}\right),\\
&w_{{\scriptscriptstyle\text{AOA}}_i}=\dfrac{1}{\sigma_{{\scriptscriptstyle\text{AOA}}_i}^2}\left(1-\frac{e_{{\scriptscriptstyle\text{AOA}}_i}^2}{\sum\limits_{i=1}^{N} e_{{\scriptscriptstyle\text{AOA}}_i}^2}\right),    
     \end{array}
 \end{equation}
where $w_{{\scriptscriptstyle\text{RSS}}_i},w_{{\scriptscriptstyle\text{TOA}}_i},w_{{\scriptscriptstyle\text{TDOA}}_i}$ and $w_{{\scriptscriptstyle\text{AOA}}_i}$ are the weights for RSS, TOA, TDOA and AOA measurements, respectively and
\begin{equation}
     \begin{array}{ll}
      e_{{\scriptscriptstyle\text{RSS}}_i}={\sigma_{{\scriptscriptstyle\text{RSS}}_i}d_i},\,
    e_{{\scriptscriptstyle\text{TOA}}_i}=\sigma_{{\scriptscriptstyle\text{TOA}}_i},\\
    e_{{\scriptscriptstyle\text{TDOA}}_i}=\sigma_{{\scriptscriptstyle\text{TDOA}}_i},\,
    e_{{\scriptscriptstyle\text{AOA}}_i}=\sigma_{{\scriptscriptstyle\text{AOA}}_i}d_i,    
     \end{array}
 \end{equation}
which denote the error impact of the measurements and these error impacts again depend on the range and noise variance of measurements.
The proposed weighing scheme in \eqref{wts} assigns higher weights to those measurements having low error impact and vice-versa. For instance, under the case of equal noise variance, RSS and AOA based measurements are given higher preference over short ranges, while TOA and TDOA based measurements are preferred over long ranges. It can be seen from \eqref{wts} that to choose weights, one needs to know the true noise variances and the actual ranges, which are obviously not known. However, one can always use a rough estimate\footnote{The noise variances can also be jointly estimated via the MLE (in an iterative manner), however we will not explore this in this work.} of the noise variances (via the sample variance estimator) and the noisy range observations to calculate the weights.

\subsection{Majorization-Minimization (MM) framework}
    The MM procedure is an iterative approach which at every iteration constructs a surrogate function that tightly upper bounds the original objective function and minimizes the surrogate function. The surrogate is constructed in such
a way that it tightly upper bounds the original function at some given point, and later the surrogate function constructed is minimized to obtain the next iterate. Let us consider an objective function $f(\bx)$ to be minimized. First, one needs to come up with a surrogate function $g(\bx|\bx^{(t)})$ that tightly upper bounds the objective $f(\bx)$ at a point $\bx^{(t)}$ such that
\begin{equation}\label{mm1}
    \begin{array}{ll}
        g(\bx|\bx^{(t)}) \geq f(\bx),  \\
        g(\bx^{(t)}|\bx^{(t)})  = f(\bx^{(t)}).
    \end{array}
\end{equation}
The minimization of this surrogate function will generate the next iterate, i.e.,
\begin{equation}\label{mm2}
   \bx^{(t+1)} =\arg\min\limits_{\bx} g(\bx|\bx^{(t)}).
\end{equation}
When these two steps are repeated multiple times,  the objective function gets monotonically decreased at each iteration, which can be shown as below:
\begin{equation}\label{mm3}
 f(\bx^{(t+1)})  \leq g(\bx^{(t+1)}|\bx^{(t)})  \leq g(\bx^{(t)}|\bx^{(t)}) =f(\bx^{(t)}).  
\end{equation}
The iterative steps of MM are repeated till convergence to get a minimizer of $f(\bx)$. The complexity and accuracy of this optimization approach depend on the choice of the surrogate function.  Interested readers can find different methods for constructing the surrogate function and applications of the MM algorithm in different fields of research in \cite{MMalgo}. In the following  subsection, we have used the aforementioned procedure of MM to design an iterative algorithm to solve our problem of interest.
\subsection{The Proposed algorithm}
Using weights proposed in \eqref{wts}, the problem of interest  can be officially stated as
\begin{equation}\label{wls}
     \begin{array}{ll}
      \hat{\bs}=\arg\min\limits_{\bs}\sum\limits_{i=1}^{N}\left\lbrace w_{{\scriptscriptstyle\text{RSS}}_i}f_{{\scriptscriptstyle\text{RSS}}_i}(\bs)
      +w_{{\scriptscriptstyle\text{TOA}}_i}f_{{\scriptscriptstyle\text{TOA}}_i}(\bs)\right.\\\hspace{2cm}\left.+w_{{\scriptscriptstyle\text{AOA}}_i}f_{{\scriptscriptstyle\text{AOA}}_i}(\bs)\right\rbrace +\sum\limits_{i=2}^{N}\left\lbrace w_{{\scriptscriptstyle\text{TDOA}}_i}f_{{\scriptscriptstyle\text{TDOA}}_i}(\bs)\right\rbrace.    
     \end{array}
 \end{equation}

As mentioned before, the problem in \eqref{wls} is non-convex, and to tackle it, we derive here an iterative algorithm based on the MM framework.  Before we develop the algorithm, we first present the following Lemmas, which will be helpful to derive the proposed algorithm. 
\setcounter{theorem}{0}
\begin{lem} \label{lemma1}
For any given $\bs^{(t)}$, the function $-{\|\bs-\bm_{i}\|}$  can be upper  bounded as 
\begin{equation}\label{eq_lemma1}
\begin{array}{ll}
  - {\|\bs-\bm_{i}\|}  \leq - \dfrac{{\left(\bs-\bm_{i}\right)^{T}}\left(\bs^{(t)}-\bm_{i}\right)}{{\|\bs^{(t)}-\bm_{i}\|}},
\end{array}
\end{equation}
and the upper bound is linear and differentiable in $\bs$.
\end{lem}
\begin{proof}
Utilizing Cauchy-Schwarz  inequality we have:
\begin{equation}
     \begin{array}{ll}
      &{\left(\bs-\bm_{i}\right)^{T}}{\left(\bs^{(t)}-\bm_{i}\right)} \leq {\|\bs-\bm_{i}\|}{\|\bs^{(t)}-\bm_{i}\|}\\
&\implies\!-{\left(\bs-\bm_{i}\right)^{T}}\!{\left(\bs^{(t)}-\bm_{i}\right)}\! \geq\! -{\|\bs-\bm_{i}\|}{\|\bs^{(t)}\!-\!\bm_{i}\|},    \end{array}
 \end{equation}
which after re-arranging gives \eqref{eq_lemma1}.
\end{proof}
\begin{lem} \label{lemma2}
For any given $\bs^{(t)}$, the function ${\|\bs\|}$  can be upper  bounded as 
\begin{equation}\label{eq_lemma2}
\begin{array}{ll}
 {\|\bs\|} \leq  {\|\bs^{(t)}\|} + \dfrac{{\|\bs\|}^{2}-{\|\bs^{(t)}\|}^{2}}{2{\|\bs^{(t)}\|}},
\end{array}
\end{equation}
and the upper bound is quadratic and differentiable in $\bs$.
\end{lem}
\begin{proof}
Let $l ={\|\bs\|}^2$. Therefore,
\begin{equation}
{\|\bs\|}= \sqrt{l}.    
\end{equation}
Since the square root  is a concave function in $\mathbb{R}_+$, linearizing it around given ${l^{(t)}}$ gives the following inequality
\begin{equation}
\sqrt{l} \leq \sqrt{l^{(t)}} + \dfrac{l - l^{(t)}}{2\sqrt{l^{(t)}}}.
\end{equation}
Substituting back for ${l}$ gives \eqref{eq_lemma2}. 
\end{proof}
\begin{lem} \label{lemma3}
For any given $u^{(t)}$ and $v^{(t)}$, the quadratic function $(u-v)^{2}$  can be upper  bounded as 
\begin{equation}\label{eq_lemma3}
(u-v)^{2} \leq 2 \left(u - \dfrac{u^{(t)}+v^{(t)}}{2}\right)^{2} + 2\left (v - \dfrac{u^{(t)}+v^{(t)}}{2}\right)^{2} 
\end{equation}
\end{lem}
\begin{proof}
We know that:
\begin{equation}
\begin{array}{ll}
\left(\left( u - \dfrac{u^{(t)}+v^{(t)}}{2}\right)-\left( \dfrac{u^{(t)}+v^{(t)}}{2}-v\right)\right)^{2}  \geq 0
\end{array}
\end{equation}
Then on expanding the expression on the left and re-arranging it, we get:
 \begin{equation}
     \begin{array}{ll}
       &\left(u-\frac{u^{(t)}+v^{(t)}}{2}\right)^2+\left(v-\frac{u^{(t)}+v^{(t)}}{2}\right)^2\geq 2\left( u - \frac{u^{(t)}+v^{(t)}}{2}\right)\left( \frac{u^{(t)}+v^{(t)}}{2}-v\right)\\
    \Rightarrow&2\left(u-\frac{u^{(t)}+v^{(t)}}{2}\right)^2+2\left(v-\frac{u^{(t)}+v^{(t)}}{2}\right)^2\geq\left(\left(u-\frac{u^{(t)}+v^{(t)}}{2}\right)+\left(\frac{u^{(t)}+v^{(t)}}{2}-v\right)\right)^2\\
    \Rightarrow&2\!\left(\!u-\!\frac{u^{(t)}\!+\!v^{(t)}}{2}\right)^2\!\!+\!2\!\left(\!v\!-\!\frac{u^{(t)}\!+\!v^{(t)}}{2}\!\right)^2\!\!\geq\!\left(u\!-\!v\right)^2\!\!.   
     \end{array}
 \end{equation}

\end{proof}

Now, we will deal with each and every term in \eqref{wls} separately. We first start with ${f}_{{\scriptscriptstyle\text{RSS}}_i}(\bs)$ and ${f}_{{\scriptscriptstyle\text{TOA}}_i}(\bs)$ in \eqref{wls}. From Lemma \eqref{lemma1}, the function ${f}_{{\scriptscriptstyle\text{RSS}}_i}(\bs)$ in \eqref{lik_rss} and  ${f}_{{\scriptscriptstyle\text{TOA}}_i}(\bs)$ in \eqref{lik_toa} can be upper bounded by the following surrogate functions at $(t+1)^{\text{th}}$ iteration:
\begin{equation}\label{sur_rss}
     \begin{array}{ll}
      g_{{\scriptscriptstyle\text{RSS}}_i}(\bs|\bs^{(t)})=\eta^2-2\eta\tilde{\lambda_i} \left(\bs-{\bm}_i\right)^T\boldsymbol{p}_i^{(t)}+\widetilde{\lambda}_i^2\|\bs-{\bm}_i\|^2,    
     \end{array}
 \end{equation}
 \begin{equation}\label{sur_toa}
     \begin{array}{ll}
      g_{{\scriptscriptstyle\text{TOA}}_i}(\bs|\bs^{(t)})=\tau_i^2-2\tau_i \left(\bs-{\bm}_i\right)^T\boldsymbol{p}_i^{(t)}+\|\bs-{\bm}_i\|^2,    
     \end{array}
 \end{equation}
where ${\bp}_i^{(t)}=\frac{\bs^{(t)}-{\bm}_i}{\|\bs^{(t)}-{\bm}_i\|^2}$ and $\widetilde{\lambda_i}=\eta\lambda_i$.

Using Lemma \ref{lemma3}, we can obtain a surrogate function for $f_{{\scriptscriptstyle\text{TDOA}}_i}(\bs)$ in \eqref{lik_tdoa}  as follows:
\begin{equation}\label{sur_tdoa1}
     \begin{array}{ll}
       g_{{\scriptscriptstyle\text{TDOA}}_i}(\bs|\bs^{(t)})=&2\left(\tau_{1i}+
       \|{\bs}-{\bm}_1\|-q_{i}^{(t)}\right)^2+\\&2\left(\|{\bs}-{\bm}_{i}\|-q_{i}^{(t)}\right)^2,  i=2,...,N 
     \end{array}
 \end{equation}
where $q_i^{(t)}=\frac{\tau_{1i}+\|{\bs}^{(t)}-{\bm}_1\|+\|{\bs}^{(t)}-{\bm}_{i}\|}{2}$.

Expanding \eqref{sur_tdoa1} and ignoring the constant terms, i.e. $\tau_{1i}^2,{q_{i}^{(t)}}^2,\tau_{1i}q_{i}^{(t)}$,  we get
\begin{equation}\label{sur_tdoa2}
     \begin{array}{ll}
     g_{{\scriptscriptstyle\text{TDOA}}_i}(\bs|\bs^{(t)}) =&2 \left(\|\bs-\bm_{1}\|^{2} +2\tau_{1i} \|\bs-\bm_{1}\|\right.\left. - 2q_{i}^{(t)}\|\bs-\bm_{1}\|\right)+\\
&2\left(\|\bs-\bm_{i}\|^{2}  -2q_{i}^{(t)}\|\bs-\bm_{i}\|\right) .   \end{array}
 \end{equation}

The terms $-\|\bs-\bm_{1}\|$ and $-\|\bs-\bm_{i}\|$ in \eqref{sur_tdoa2} can be once again upper bounded using Lemma \ref{lemma1} while  the term $\|\bs-\bm_{1}\|$ is upper bounded using Lemma \ref{lemma2}, and a refined surrogate function  can be obtained as follows:
\begin{equation}
     \begin{array}{ll}
      g_{{\scriptscriptstyle\text{TDOA}}_i}({\bs}|\bs^{(t)})=2\left(\|\bs-\bm_{1}\|^2+h_i^{(t)}\|\bs-\bm_{1}\|^2-2(\bs-\bm_{1})^Tq_{i}^{(t)}{\boldsymbol{p}_{1}^{(t)}}\right)
    \\\hspace{2.3cm}+ 2\left(\|\bs-\bm_{i}\|^{2} -2q_{i}^{(t)}{\boldsymbol{p}_{i}^{(t)}}^T\left(\bs-{\bm}_{i}\right)\right),    
     \end{array}
 \end{equation}
where $h_i^{(t)}=\frac{\tau_{1i}}{\|\bs^{(t)}-\bm_{1}\|}$.

The surrogate function for function $f_{{\scriptscriptstyle\text{AOA}}_i}({\bs})$ in (\ref{lik_aoa}) can be constructed  using Lemma \ref{lemma3} by upper bounding the second term $(\bk^{T}(\bs -\bm_i)- \|\bs-\bm_{i}\|\cos \theta_{i})\!^2$ as:
\begin{equation}\label{sur_aoa1}
     \begin{array}{ll}
     g_{{\scriptscriptstyle\text{AOA}}_i}({\bs}|\bs^{(t)})=&\hspace{-3mm} \left(\bc_{i}^{T}(\bs -\bm_i) \right)^{2}+2\left (\left(\bk^{T}(\bs -\bm_i)- u_{i}^{(t)}\right)^{2} \right)\\
&\hspace{-3mm}  +2\left( \left(\|\bs-\bm_{i}\|\cos \theta_{i}-u_{i}^{(t)}\right)^{2}\right),     
     \end{array}
 \end{equation}
where $u_{i}^{(t)} = \dfrac{\bk^{T}(\bs^{t}-\bm_{i}) + \|\bs^{t}-\bm_{i}\|\cos \theta_{i}}{2}$.

On expanding the surrogate function in \eqref{sur_aoa1} and ignoring the constant terms we get
\begin{equation}
     \begin{array}{ll}
      g_{{\scriptscriptstyle\text{AOA}}_i}({\bs}|\bs^{(t)})=&\hspace{-3mm} \left(\bs^{T}\bc_{i}\bc_{i}^{T}\bs -2 \bs^{T}\bc_{i}\bc_{i}^{T}\bm_{i}\right) +2\big\lbrace\left(\bs-\bm_i\right)^{T}\bk\bk^T+\left(\bs-\bm_{i}\right)\\
&\hspace{-3mm}- 2u_{i}^{(t)}\bk^{T}(\bs-\bm_{i})
+ \|\bs-\bm_{i}\|^{2}\cos^{2}\theta_{i}-2u_{i}^{(t)}\|\bs-\bm_{i}\|\cos\theta_{i}\big\rbrace,     
     \end{array}
 \end{equation}
which can be rewritten as:
 \begin{equation}\label{sur_aoa2}
     \begin{array}{ll}
    g_{{\scriptscriptstyle\text{AOA}}_i}({\bs}|\bs^{(t)})=&\hspace{-3mm} \left(\bs^{T}(\bc_{i}\bc_{i}^{T}-\bI)\bs +\|\bs\|^2-2 \bs^{T}\bc_{i}\bc_{i}^{T}\bm_{i}\right)+\\
&\hspace{-3mm}2\bigg\lbrace\left(\bk^T\left(\bs-\bm_{i}\right)\right)^{2}- 2u_{i}^{(t)}\bk^T\left(\bs-\bm_{i}\right)+\\&\hspace{-3mm} \|\bs\|^2\cos^{2}\theta_{i}-2\bs^{T}\bm_{i}\cos^{2}\theta_{i} -2u_{i}^{(t)}\|\bs-\bm_{i}\|\cos\theta_{i}\bigg\rbrace.      \end{array}
 \end{equation}
 Since $\bs^{T}(\bc_{i}\bc_{i}^{T}-\bI)\bs$ is a concave function in $\bs$, we can majorize it by linearizing it at some given $\bs^{(t)}$ as follows:
\begin{equation}
     \begin{array}{ll}
      \bs^{T}(\bc_{i}\bc_{i}^{T}-\bI)\bs\leq2\left(\left(\bc_{i}\bc_{i}^{T}-\bI\right)\bs^{(t)}\right)^{T}(\bs- \bs^{(t)})+\text{const}.
     \end{array}
 \end{equation}
Also, in \eqref{sur_aoa2}, depending on the sign of
$u_{i}^{(t)}\cos\theta_{i}$, the surrogate in \eqref{sur_aoa2} can be further upper bounded via  two different surrogate functions, i.e., when $u_{i}^{(t)}\cos\theta_{i} >0$, it can be majorized using Lemma \ref{lemma1} and when $u_{i}^{(t)}\cos\theta_{i} <0$, it can be majorized using Lemma \ref{lemma2}. 
More clearly, the last term in \eqref{sur_aoa2}, i.e.  $-2u_{i}^{(t)}\cos\theta_{i}\|\bs-\bm_{i}\|$ can be written as
\begin{equation}\label{sgn_t}
     \begin{array}{ll}
      \hspace{-1mm}-2u_{i}^{(t)}\cos\theta_{i}\|\bs-\bm_{i}\|=&\hspace{-3mm}-2u_{i}^{(t)}\cos\theta_{i}\frac{\text{sgn}(u_{i}^{(t)}\cos\theta_{i})+1}{2}\|\bs-\bm_{i}\|\\
\hspace{-1mm}&\hspace{-3mm}+2u_{i}^{(t)}\cos\theta_{i}\frac{\text{sgn}(u_{i}^{(t)}\cos\theta_{i})-1}{2}\|\bs-\bm_{i}\|.   
     \end{array}
 \end{equation}
 So, the term $\dfrac{\text{sgn}(u_{i}^{(t)}\cos\theta_{i}))+1}{2}$ gets activated when $u_{i}^{(t)}\cos\theta_{i} >0$ and the term $\dfrac{\text{sgn}(u_{i}^{(t)}\cos\theta_{i})-1}{2}$ gets activated when $u_{i}^{(t)}\cos\theta_{i} <0$. We now majorize  both the terms in \eqref{sgn_t} using Lemma \ref{lemma1} and Lemma \ref{lemma2} as follows:
\begin{equation}
     \begin{array}{ll}
      -2u_{i}^{(t)}\cos\theta_{i}\|\bs-\bm_{i}\|\leq&-2u_{i}^{(t)}\cos\theta_{i}\frac{\text{sgn}(u_{i}^{(t)}\cos\theta_{i})+1}{2} (\bs-\bm_{i})^{T}\frac{(\bs^{(t)}-\bm_{i})}{\|\bs^{(t)}-\bm_{i}\|}
 \\&+u_{i}^{(t)}\cos\theta_{i}\frac{\text{sgn}(u_{i}^{(t)}\cos\theta_{i})-1}{2}\frac{\|\bs-\bm_{i}\|^{2}} {\|\bs^{(t)}-\bm_{i}\|}+\text{const}.    \end{array}
 \end{equation}
Finally, the surrogate function for  ${f}_{{\scriptscriptstyle\text{AOA}}_i}(\bs)$ can be simplified as:
\begin{equation}\label{sur_aoa3}
     \begin{array}{ll}
      g_{{\scriptscriptstyle\text{AOA}}_i}({\bs}|\bs^{(t)})\!= &\left(2\bomega_{i}^{T}\bs +\|\bs\|^2 \right)+2\left(\bk^T\left(\bs-\bm_{i}\right)\right)^{2}+\\
&2\Big\lbrace- 2u_{i}^{(t)}\bk^T\left(\bs-\bm_{i}\right)+ \|\bs\|^2\cos^{2}\theta_{i}\\
& -2\bs^{T}\bm_{i}\cos^{2}\theta_{i}+\Omega_i \boldsymbol{p}_i^{(t)}(\bs-\bm_{i})^{T} +\rho_i{\|\bs-\bm_{i}\|^{2}} \Big\rbrace ,   
     \end{array}
 \end{equation}
where $\bomega_{i} =\left(\left(\bc_{i}\bc_{i}^{T}-\bI\right)\bs^{(t)}-\bc_{i}\bc_{i}^{T}\bm_{i}\right)$ and
\begin{equation*}
     \begin{array}{ll}
       &\Omega_i=-2u_{i}^{(t)}\left(\frac{\text{sgn}(u_{i}^{(t)}\cos\theta_{i})+1}{2}\right)\cos\theta_{i},  
     \end{array}
 \end{equation*}
\begin{equation*}
     \begin{array}{ll}
    &\rho_i=u_{i}^{(t)}\left(\frac{\text{sgn}(u_{i}^{(t)}\cos\theta_{i})-1}{2{\|\bs^{(t)}-\bm_{i}\|}}\right)\cos\theta_{i}.   
     \end{array}
 \end{equation*}

Using \cref{sur_rss,sur_toa,sur_tdoa2,sur_aoa3}, the final surrogate function for the objective function in \eqref{wls} at $(t+1)^{\text{th}}$ iteration can be written as \begin{equation}\label{sur_hyb}
     \begin{array}{ll}
       g({\bs}|\bs^{(t)})=&\sum\limits_{i=1}^{N}\big\lbrace w_{{\scriptscriptstyle\text{RSS}}_i}g_{{\scriptscriptstyle\text{RSS}}_i}({\bs}|\bs^{(t)})+w_{{\scriptscriptstyle\text{TOA}}_i}g_{{\scriptscriptstyle\text{TOA}}_i}({\bs}|\bs^{(t)})+\\& w_{{\scriptscriptstyle\text{AOA}}_i}g_{{\scriptscriptstyle\text{AOA}}_i}({\bs}|\bs^{(t)})\big\rbrace  + \sum\limits_{i=2}^{N}\big\lbrace w_{{\scriptscriptstyle\text{TDOA}}_i}g_{{\scriptscriptstyle\text{TDOA}}_i}({\bs}|\bs^{(t)})\big\rbrace.    
     \end{array}
 \end{equation}
 
Substituting for the surrogates in \eqref{sur_hyb} and ignoring the constant terms, we get:
\begin{equation}
     \begin{array}{ll}
      g({\bs}|\bs^{(t)})=\bs^T{\bZ}^{(t)}\bs-2{(\by^{(t)})}^T{\bs},    
     \end{array}
 \end{equation}
where $\bZ^{(t)}=\text{diag}(\left[z,z,z+\tilde{z}\right])$ and 
\begin{equation}\label{z1}
     \begin{array}{ll}
       z\triangleq&\sum\limits_{i=1}^N \Big\lbrace w_{{\scriptscriptstyle\text{RSS}}_i}\widetilde{\lambda}_i^2+w_{{\scriptscriptstyle\text{TOA}}_i}+w_{{\scriptscriptstyle\text{AOA}}_i}\left(1+2\cos^2\theta_i+2\rho_i\right)\Big\rbrace \\
       &\quad\quad +  \sum\limits_{i=2}^{N} \Big\lbrace  2w_{{\scriptscriptstyle\text{TDOA}}_i}\left(2+h_i^{(t)}\right) \Big\rbrace, 
     \end{array}
 \end{equation}
 \begin{equation}\label{z2}
     \begin{array}{ll}
      \tilde{z}\triangleq\sum\limits_{i=1}^N 2w_{{\scriptscriptstyle\text{AOA}}_i},    
     \end{array}
 \end{equation}
 \begin{equation}\label{z3}
     \begin{array}{ll}
       {(\by^{(t)})}^T\triangleq&\sum\limits_{i=1}^N \bigg\lbrace w_{{\scriptscriptstyle\text{RSS}}_i}\left(\eta\widetilde{\lambda}_i({\bp_i^{(t)}})^T+\widetilde{\lambda}_{i}^2\bm_i^T\right)+\\
    &\quad\quad w_{{\scriptscriptstyle\text{TOA}}_i}\left(\tau_i({\bp_i^{(t)}})^T+\bm_i^T\right)+\\
    &\quad\quad w_{{\scriptscriptstyle\text{AOA}}_i}\bigg(-\bomega_i^T+2u_i^{(t)}\bk^T+2\bm_i^T\bk\bk^T+\\&\quad\quad\quad 2\cos^2\theta_i\bm_i^T-\Omega_i(\bp_i^{(t)})^T+2\rho_i\bm_i^T\bigg)\bigg\rbrace+\\
    &\sum\limits_{i=2}^{N}   w_{{\scriptscriptstyle\text{TDOA}}_i}\left((2\bm_{i+1}^T+2\bm_{1}^T+\right.\\&\quad\quad\left.2q_{i}^{(t)}(\bp_{i}^{(t)})^T+2q_{i}^{(t)}(\bp_{1}^{(t)})^T+2h_i^{(t)}m_1^T\right).
     \end{array}
 \end{equation}

So, at $(t+1)^{\text{th}}$ iteration, the problem of minimizing surrogate function is given by:
\begin{equation}\label{s_t}
     \begin{array}{ll}
       \bs^{(t+1)}=\arg\min\limits_{\bs}\bs^T{\bZ}^{(t)}\bs-2({\by^{(t)}})^T{\bs},   
     \end{array}
 \end{equation}
which has a simple closed-form solution given by:
\begin{equation}\label{sol_fin}
     \begin{array}{ll}
        \bs^{(t+1)}={\by}^{(t)}\odot [z^{-1}, z^{-1},(z+\tilde{z})^{-1} ]^T.   
     \end{array}
 \end{equation}

Now,  solving \eqref{sol_fin} in an iterative manner for $T^{\text{max}}$ iterations or until some convergence threshold ($\epsilon_c$) is reached, provides a  minimizer of \eqref{wls}. The algorithm framework can also work for  standalone  measurements or any limited combination of measurements (TOA-RSS, RSS-AOA, etc). For instance, to solve RSS based localization, we can set weights of RSS, i.e. $w_{{\scriptscriptstyle\text{RSS}}_i}$ as in \eqref{wts} and set others to 0 and solve \eqref{sol_fin}, while for hybrid RSS-TOA localization, we can set weights for $w_{{\scriptscriptstyle\text{RSS}}_i}$ and $w_{{\scriptscriptstyle\text{TOA}}_i}$ as in \eqref{wts} and set $w_{{\scriptscriptstyle\text{TDOA}}_i}=0$ and $w_{{\scriptscriptstyle\text{AOA}}_i}=0$. The pseudo-code for the proposed hybrid method is given in Algorithm \ref{algo1}. 
\begin{algorithm}[b!]
\caption{Proposed Hybrid Localization algorithm based on the Majorization-Minorization (MM) framework.}
\textbf{Input}: $\bs^{(0)},\bm_i,\forall i\in\mathbb{N}_n,T^{\text{max}}=1000$ and $\epsilon_c=10^{-3}$
\begin{enumerate}
    \item Initialize $t=0$
    \item Calculate weights as in \eqref{wts}. 
    \item \textbf{While} {$\frac{\|\bs^{(t+1)}-\bs^{(t)}\|}{\|\bs^{(t)}\|}\geq\epsilon_c$ or $t\leq T^{\text{max}}$}
    \item Compute $\bZ^{(t)}$ and $\by^{(t)}$ using \eqref{z1}, \eqref{z2} and \eqref{z3}. 
    \item Solve $\bs^{(t+1)}$ using \eqref{sol_fin}.
\item $t=t+1$
\item \textbf{end while}
\end{enumerate}
% \begin{algorithmic}
% \State Initialize $t=0$
% \State Calculate weights as in \eqref{wts}. 
% \While {$\frac{\|\bs^{(t+1)}-\bs^{(t)}\|}{\|\bs^{(t)}\|}\geq\epsilon_c$ or $t\leq T^{\text{max}}$}
% \State Compute $\bZ^{(t)}$ and $\by^{(t)}$ using \eqref{z1}, \eqref{z2} and \eqref{z3}. 
% \State Solve $\bs^{(t+1)}$ using \eqref{sol_fin}.
% \State $t=t+1$
% \EndWhile
% \end{algorithmic}
\textbf{Output}: $\hat{\bs}=\bs^{(t)}$
\label{algo1}
\end{algorithm}

\subsection{Computational complexity and convergence of the proposed algorithm}
We first discuss the computational complexity of our proposed method. In each iteration, we require the computation of the scalars $z$ and $\tilde{z}$ (which has total complexity of $\mathcal{O}(Nn)$) and vector $\by$ (which has total complexity of $\mathcal{O}(Nn^2)$). So, the per iteration  complexity of the proposed algorithm is $\mathcal{O}(Nn(n+1))$. Since $N>>n$ in practical applications, the overall complexity of the proposed algorithm is  $\mathcal{O}(Nn^2)$.

Next, we discuss the proof of convergence of the proposed algorithm. Since the proposed algorithm is based on the MM approach, the iterates generated by our algorithm  will monotonically decrease the objective function (which is clear from \eqref{mm3}). Now, as the objective function given in \eqref{wls} is bounded below by zero, these iterative updates will  converge to a limit point. The following Lemma   establishes that every limit point generated by the proposed algorithm is a stationary point of \eqref{wls}. 
\begin{lem}
Every limit point generated by the proposed algorithm is the stationary point of \eqref{wls}.
 \end{lem}
  \begin{proof}
Let us assume that a sub-sequence $\{\bs^{x_j}\}$ generated by the proposed algorithm converge to a limit point $\bs^\infty$, then from \eqref{mm1} and \eqref{mm2} we can write 
\begin{equation}
\begin{array}{ll}
 g(\bs^{x_{j+1}}|\bs^{x_{j+1}}) = f(\bs^{x_{j+1}}) \leq f(\bs^{x_{j}+1}) &\hspace{-2mm}\leq g(\bs^{x_{j}+1}|\bs^{x_{j}})\\ &\hspace{-2mm}\leq g(\bs|\bs^{x_{j}}).   
\end{array}
\end{equation} 
Taking \( j\rightarrow \infty\) gives
\[g(\bs^\infty|\bs^\infty)\leq g(\bs|\bs^\infty),\] which further gives
\[g'(\bs|\bs^\infty;d)|_{\bs=\bs^\infty} = 0.\]
 where $g'(\bs;\bd)$ is the first order directional derivative of $g(\bs)$ at point $\bx$ in direction $\bd$. Since \(g'(\bs|\bs^{(t)};d)|_{\bs=\bs^{(t)}} = f'(\bs^{(t)};\bd)\) we get $f'(\bs^\infty;\bd) = 0$.  Hence $\bs^\infty$ is a stationary point of the objective function.
\end{proof}

\section{Simulation Results}\label{sec_Sim}
This section presents the numerical performance of the proposed method under different parameter settings obtained via Monte Carlo simulations. The number of Monte-Carlo runs $(M_c)$ performed in the simulation studies is  1000. The root mean square error (RMSE) is considered the main performance metric for determining localization error which is calculated as:
\begin{align}
\text{RMSE}=\sqrt{\frac{1}{M_c}\sum_{j=1}^{M_c}\left\lvert\left\lvert{\bs}-\hat{{\bs}}_j\right\rvert\right\rvert^2},
\end{align}
where ${\bs}$ and $\hat{{\bs}}_j$ are the true and estimated value of the location of the source node in the $j^{\text{th}}$ Monte-Carlo run. 
The generalized parameter setting for the network simulation is considered as follows. We consider a 3D network where a source node and a set of $N$ anchor nodes are randomly deployed on a spherical surface of radius $R$ meters.  For the sake of simplicity, we assume  that the measurement noises in RSS, AOA, and  TOA/TDOA at different anchors are distributed with equal variance such that $\sigma_{{\scriptscriptstyle\text{AOA}}_i}^2=\sigma_{\scriptscriptstyle\text{AOA}}^2,\sigma^2_{{\scriptscriptstyle\text{RSS}}_i}=\sigma_{\scriptscriptstyle\text{RSS}}^2$,$\sigma^2_{{\scriptscriptstyle\text{TDOA}}_i}=\sigma_{\scriptscriptstyle\text{TDOA}}^2$ and $\sigma^2_{{\scriptscriptstyle\text{TOA}}_i}=\sigma_{\scriptscriptstyle\text{TOA}}^2,\forall i\in\mathbb{N}_n$. Note that the standard deviation  of the noise, i.e.  $\sigma_{\scriptscriptstyle\text{AOA}},\sigma_{\scriptscriptstyle\text{RSS}}$ and $\sigma_{\scriptscriptstyle\text{TOA}}$ are denoted in $^{\circ}$, dB and meters, respectively. The RSS parameters used in the simulations are ${L_o}=20$ dB and  $\gamma=2.5$.  

Particularly, we analyze the performance behavior of  the proposed MM algorithm  with respect to  varying noise levels, the number of anchors, convergence behavior, deployment range, and the effect of NLOS errors. Note that the proposed framework can provide several localization methods by fusing multiple heterogeneous measurements, and these localization methods are referred to as their initials. Like, the standalone AOA and TDOA based localization methods and hybrid TDOA-AOA based localization methods are referred as A, D and DA, respectively. By utilizing the combination of AOA, RSS, TOA, and TDOA measurements, we analyze 11 different  methods based on the proposed algorithm, i.e. 4 hybrid range methods and 7 hybrid range-angle methods. Hence, this section provides a  complete analysis of all possible hybrid localization  methods w.r.t. various network scenarios. The numerical performance of the proposed algorithm is compared with the most commonly used WLS approach, which is nothing but the conventional least squares approach with weights from \eqref{wts} (combination used for comparison is TDRA, named as TDRA$^{\text{W}}$), RSS-AOA [26] (named as RA-[26]), RSS-TOA [15] (named as TR-SDP) and TDOA-AOA [23] (named as DA-SDR). The initial estimate of the source location \eqref{s_t} for the proposed algorithm is randomly chosen with values taken from the uniform distribution $\mathcal{U}(0,R/4)$. The noise variances corresponding to different types of measurements used in the weights are obtained via the sample variance estimator.

%\subsection{Proposed Hybrid Localization Framework}
We divide our numerical simulations into five subsections. First, we derive the CRLB for the hybrid measurements which is used as a benchmark in the performance analysis. In the second part, the convergence  of the proposed method is shown numerically. Along with this, the computational complexity of the proposed method is also analyzed in terms of average computational time. Next, the impacts of noise, deployment area, and the number of anchors on the localization accuracy of the proposed algorithm are shown. 
\subsection{CRLB of the hybrid model}
In this subsection, the CRLB for source location estimate is calculated for the hybrid TOA-TDOA-RSS-AOA measurement model.  The CRLB matrix is defined as the inverse of the Fisher Information matrix (FIM). 
% \begin{equation}\label{crlb_s}
%     \begin{array}{ll}
%         \text{CRLB}=\text{FIM}^{-1}.
%     \end{array}
% \end{equation}
For the data models in \cref{rss1,toa,tdoa,aoa1,aoa2}, the corresponding model specific FIM can be expressed as: 
\begin{equation}
    \begin{array}{ll}
    \text{FIM}_{\scriptscriptstyle\text{TOA}}
         =\textbf{J}_{\scriptscriptstyle\text{TOA}}^T\mathbf{\Sigma}^{-1}_{\scriptscriptstyle\text{TOA}}\textbf{J}_{\scriptscriptstyle\text{TOA}},\;
         \text{FIM}_{\scriptscriptstyle\text{RSS}}
         =\textbf{J}_{\scriptscriptstyle\text{RSS}}^T\mathbf{\Sigma}^{-1}_{\scriptscriptstyle\text{RSS}}\textbf{J}_{\scriptscriptstyle\text{RSS}},\\
         \text{FIM}_{\scriptscriptstyle\text{TDOA}}
         =\textbf{J}{\scriptscriptstyle\text{TDOA}}^T\mathbf{\Sigma}^{-1}{\scriptscriptstyle\text{TDOA}}\textbf{J}{\scriptscriptstyle\text{TDOA}},\\
         \text{FIM}_{{\scriptscriptstyle\text{AOA}}_\phi}
         =\textbf{J}_{{\scriptscriptstyle\text{AOA}}_\phi}^T\mathbf{\Sigma}^{-1}_{\scriptscriptstyle\text{AOA}}\textbf{J}_{{\scriptscriptstyle\text{AOA}}_\phi},\;
         \text{FIM}_{{\scriptscriptstyle\text{AOA}}_\theta}
         =\textbf{J}_{{{\scriptscriptstyle\text{AOA}}_\theta}}^T\mathbf{\Sigma}^{-1}_{\scriptscriptstyle\text{AOA}}\textbf{J}_{{\scriptscriptstyle\text{AOA}}_\theta},
    \label{fimS}
    \end{array}
\end{equation}
where the covariance matrices are given as:
\begin{equation}
    \begin{array}{ll}
    \mathbf{\Sigma}_{\scriptscriptstyle\text{TOA}}=\text{diag} [\sigma^2_{{\scriptscriptstyle\text{TOA}}_1},....,\sigma^2_{{\scriptscriptstyle\text{TOA}}_N}],\;
    \mathbf{\Sigma}_{\scriptscriptstyle\text{RSS}}=\text{diag} [\sigma^2_{{\scriptscriptstyle\text{RSS}}_1},....,\sigma^2_{{\scriptscriptstyle\text{RSS}}_N}],\\
         \mathbf{\Sigma}{\scriptscriptstyle\text{TDOA}}=\text{diag} [\sigma^2_{{\scriptscriptstyle\text{TDOA}}_2},....,\sigma^2_{{\scriptscriptstyle\text{TDOA}}_{N}}],\\
         
         \mathbf{\Sigma}_{{\scriptscriptstyle\text{AOA}}_\phi}=\mathbf{\Sigma}_{{\scriptscriptstyle\text{AOA}}_\theta}=\text{diag} [\sigma^2_{{\scriptscriptstyle\text{AOA}}_1},....,\sigma^2_{{\scriptscriptstyle\text{AOA}}_N}],
    \label{covar}
    \end{array}
\end{equation}
\begin{figure}[!t]
    \centering
    \includegraphics[width=0.5\textwidth]{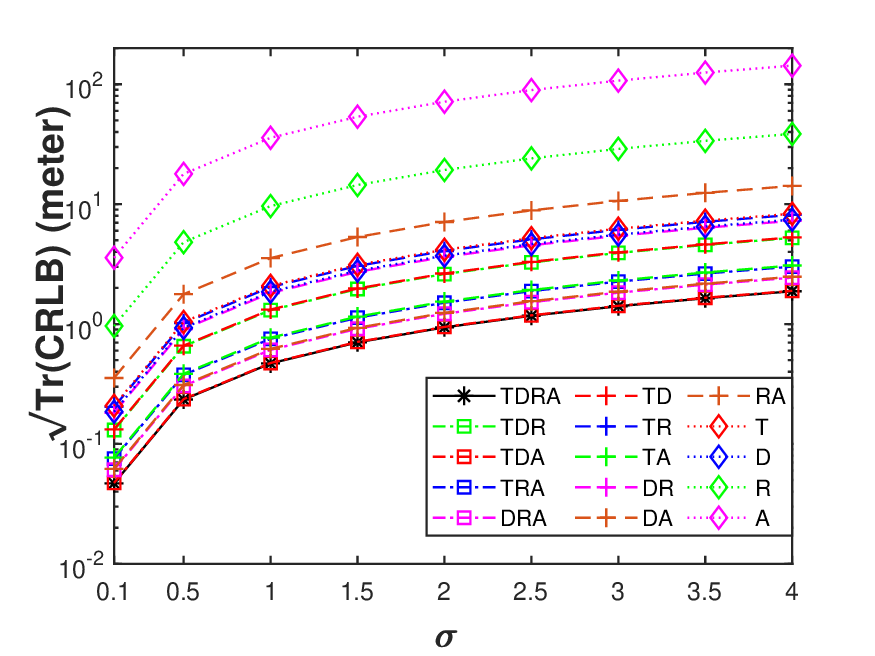}
    \caption{CRLB vs $\sigma$ for different models}
    \label{fig:crlb}
\end{figure}
and the J matrices corresponding to TOA, RSS, TDOA, $\text{AOA}_{\phi}$ and $\text{AOA}_{\theta}$ are given by:
\begin{equation}
    \begin{array}{cc}
         \textbf{J}_{\scriptscriptstyle\text{TOA}}=\begin{bmatrix}\frac{ (\bs-\bm_1)T}{\|\bs-\bm_1\|}\\
         \vdots\\
         \frac{ (\bs-\bm_N)^T}{\|\bs-\bm_N\|}
\end{bmatrix},\; 
         \textbf{J}_{\scriptscriptstyle\text{RSS}}=\eta\begin{bmatrix}\frac{ (\bs-\bm_1)^T}{\|\bs-\bm_1\|^2}\\
         \vdots\\
         \frac{ (\bs-\bm_N)^T}{\|\bs-\bm_N\|^2}
\end{bmatrix},
    \label{j_t}
    \end{array}
\end{equation}
\begin{equation}
    \begin{array}{cc}
         \textbf{J}_{{\scriptscriptstyle\text{TDOA}}}=\begin{bmatrix}\frac{(\bs-\bm_{2})^T}{\|\bs-\bm_{2}\|}-\frac{ (\bs-\bm_{1})^T}{\|\bs-\bm_{1}\|}\\
         \vdots\\
         \frac{ (\bs-\bm_{N})^T}{\|\bs-\bm_{N}\|}-\frac{ (\bs-\bm_{1})^T}{\|\bs-\bm_{1}\|}
\end{bmatrix},
    \end{array}
\end{equation}
\begin{equation}
    \begin{array}{cc}
         \textbf{J}_{{\scriptscriptstyle\text{AOA}}_{\phi}}\footnotemark=\begin{bmatrix}\frac{\sin\phi_1}{d_{xy1}}&-\frac{\cos\phi_1}{d_{xy1}}&0\\
         \vdots&\vdots&\vdots\\
         \frac{\sin\phi_N}{d_{xyN}}&-\frac{\cos\phi_N}{d_{xyN}}&0
\end{bmatrix}, 
\text{ and}\\ \textbf{J}_{{\scriptscriptstyle\text{AOA}}_{\theta}}=\begin{bmatrix}\frac{\cos\phi_1\cos\theta_1}{\|\bs-\bm_1\|}&\frac{\sin\phi_1\cos\theta_1}{\|\bs-\bm_1\|}&-\frac{\sin\theta_1}{\|\bs-\bm_1\|}\\
         \vdots&\vdots&\vdots\\
        \frac{\cos\phi_N\cos\theta_N}{\|\bs-\bm_N\|}&\frac{\sin\phi_N\cos\theta_N}{\|\bs-\bm_N\|}&-\frac{\sin\theta_N}{\|\bs-\bm_N\|}
\end{bmatrix}.
    \label{a_t}
    \end{array}
\end{equation}
\footnotetext{$d_{xyi}$ is the distance in xy-coordinates.}
For the hybrid measurement model, its FIM is given by the summation of the FIM corresponding to the individual measurement models. 
\begin{equation}
    \begin{array}{ll}
      \text{FIM}_{\scriptscriptstyle\text{Hyb}}=  \text{FIM}_{\scriptscriptstyle\text{TOA}}+\text{FIM}_{\scriptscriptstyle\text{TDOA}}+\text{FIM}_{\scriptscriptstyle\text{RSS}}+\text{FIM}_{{\scriptscriptstyle\text{AOA}}_\phi}+\text{FIM}_{{\scriptscriptstyle\text{AOA}}_{\theta}}.
    \label{fim_s}
    \end{array}
\end{equation}

The CRLBs corresponding to the different models are compared in Figure \ref{fig:crlb}. The parameters set for the simulation are as follows: \(N=5,\hspace{2mm}R=50 \text{ meters},\hspace{2mm}\sigma=\sigma_{\scriptscriptstyle\text{TOA}}=\sigma_{\scriptscriptstyle\text{TDOA}}=\sigma_{\scriptscriptstyle\text{RSS}}=\sigma_{\scriptscriptstyle\text{AOA}}\). From the figure, it is seen that the CRLB values corresponding to the different models increase  with the increase in the standard deviation of the noise, as expected. It can be noticed from the plot that the CRLB of the TDRA  is the best among all the models.

\subsection{Convergence analysis}
\begin{figure}[ht!]
\begin{subfigure}[t]{0.24\textwidth}\centering
 \includegraphics[width=0.8\textwidth]{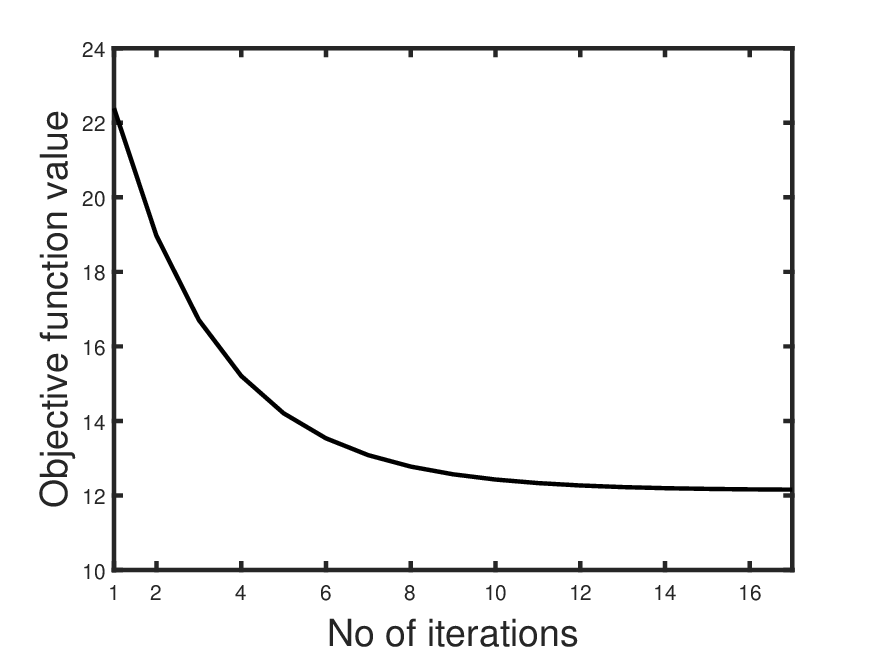}
 \caption{TOA-TDOA-RSS-AOA }
 \label{fig_iter1}
\end{subfigure}
\begin{subfigure}[t]{0.24\textwidth}\centering
 \includegraphics[width=0.8\textwidth]{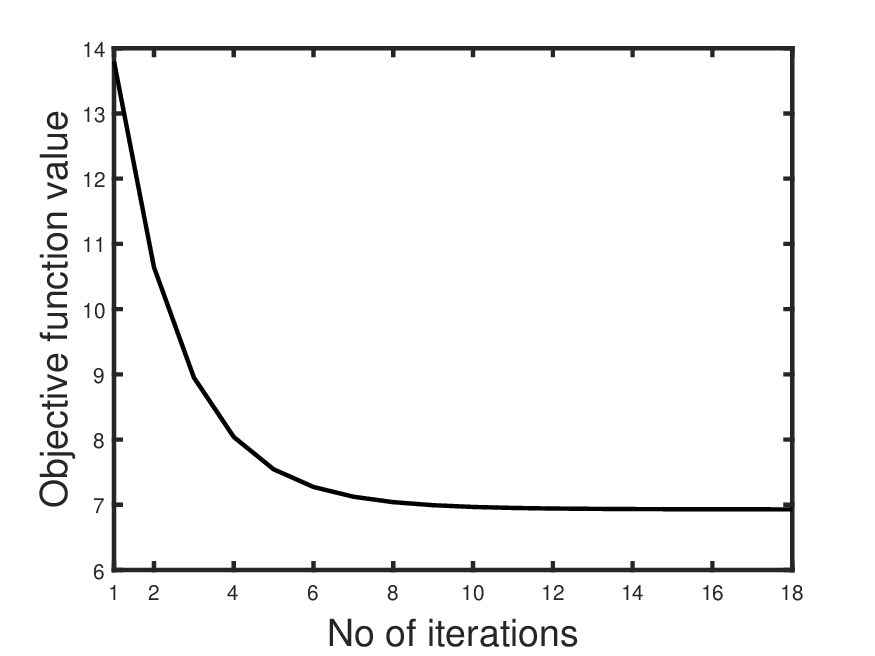}
 \caption{TOA-RSS-AOA}
 \label{fig_iter2}
\end{subfigure}
\begin{subfigure}[t]{0.24\textwidth}\centering
 \includegraphics[width=0.8\textwidth]{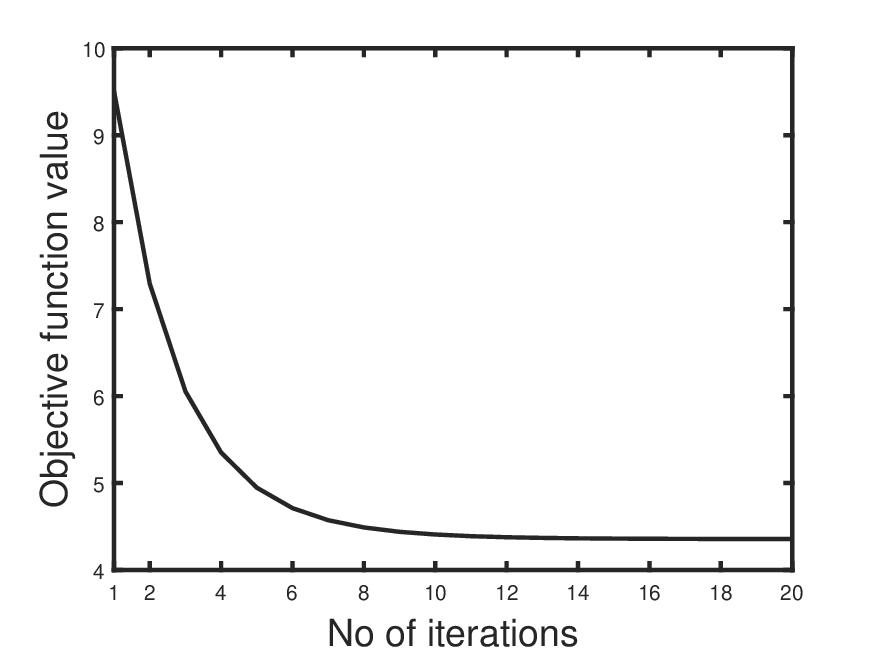}
 \caption{TOA-AOA}
 \label{fig_iter3}
\end{subfigure}
\begin{subfigure}[t]{0.24\textwidth}\centering
 \includegraphics[width=0.8\textwidth]{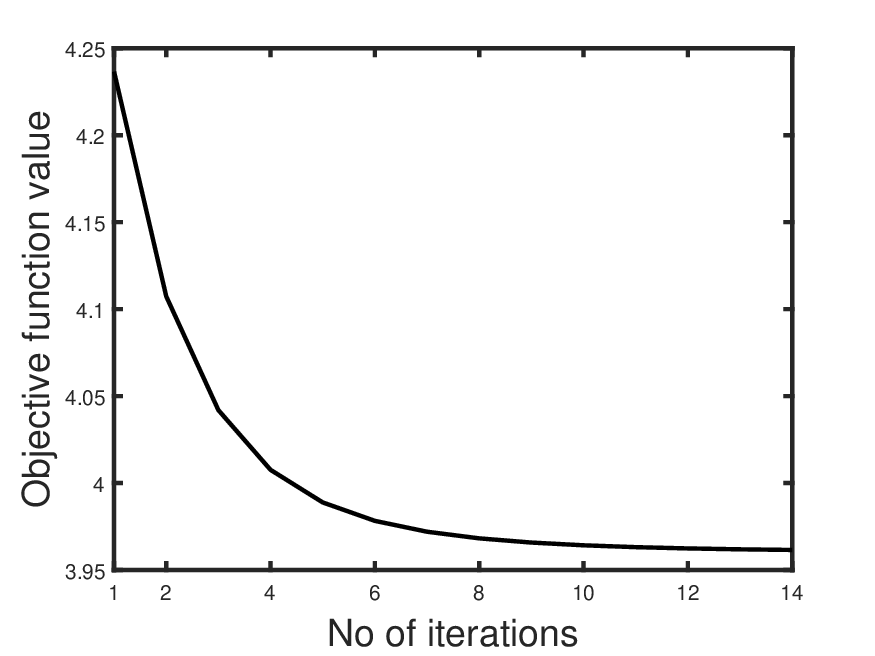}
 \caption{AOA}
 \label{fig_iter4}
\end{subfigure}
\caption{Convergence plots of the proposed algorithms corresponding to different combinations of measurements.}
\label{fig_conv}
\end{figure}
\begin{figure*}[h!]
\centering
\begin{subfigure}[t]{0.47\textwidth}\centering
 \includegraphics[width=\textwidth,height=6cm]{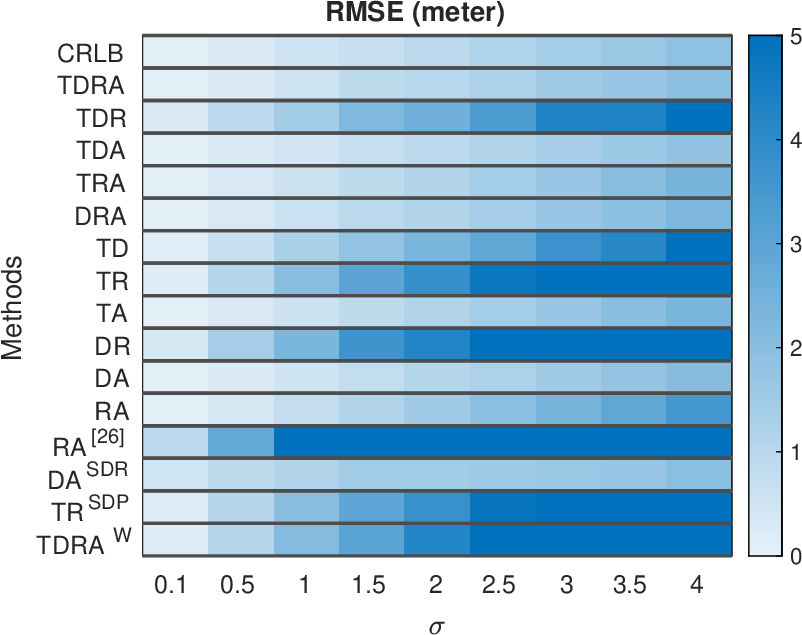}
 \caption{}
 \label{fig_all21}
\end{subfigure}
\begin{subfigure}[t]{0.47\textwidth}\centering
 \includegraphics[width=\textwidth,height=6cm]{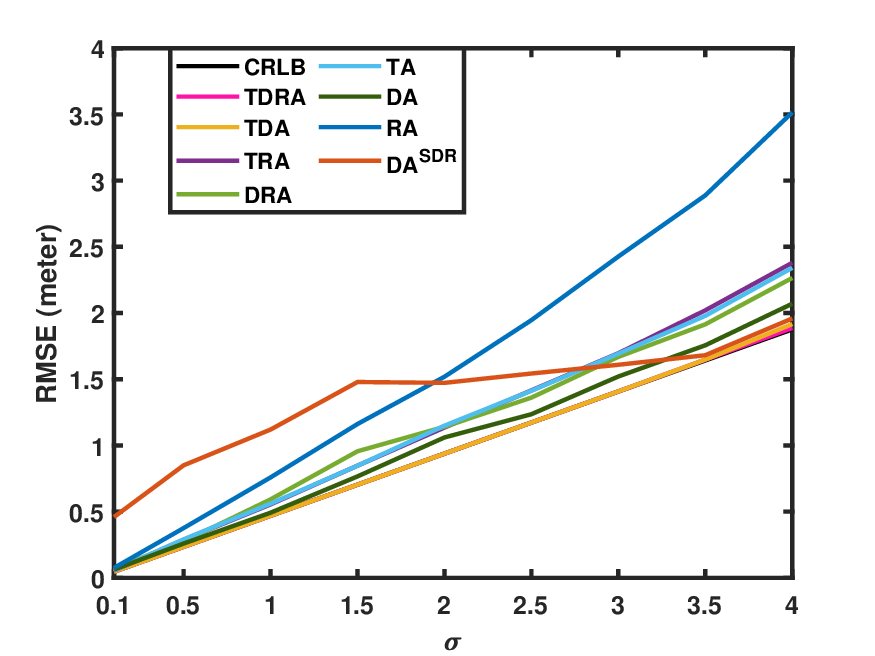}
 \caption{}
 \label{fig_all31}
\end{subfigure}
\caption{Impact of noise in all measurements on localization accuracy.}
\label{allnoise1}
\end{figure*}

\begin{figure*}[ht!]
\centering
\begin{subfigure}[t]{0.47\textwidth}\centering
 \includegraphics[width=\textwidth,height=6cm]{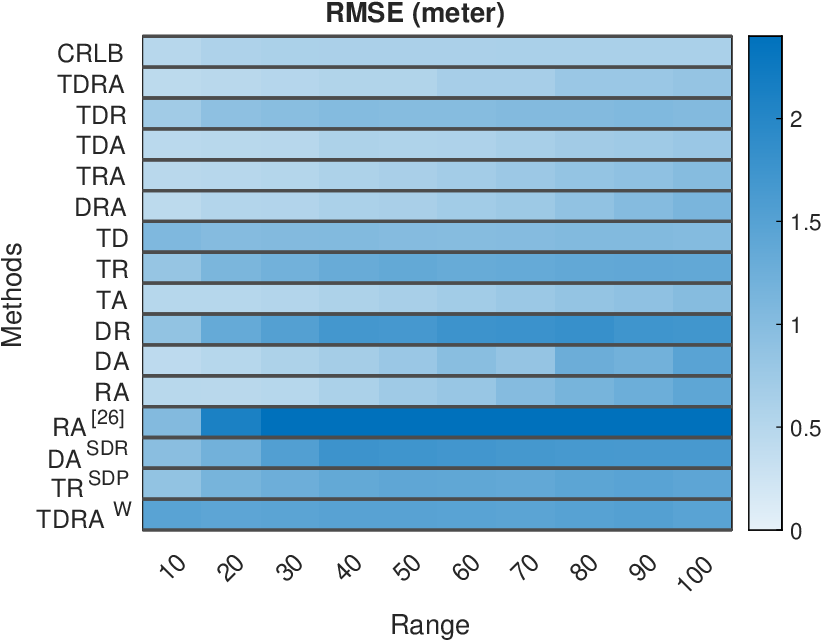}
 \caption{}
 \label{fig_rang2}
\end{subfigure}
\begin{subfigure}[t]{0.47\textwidth}\centering
 \includegraphics[width=\textwidth,height=6cm]{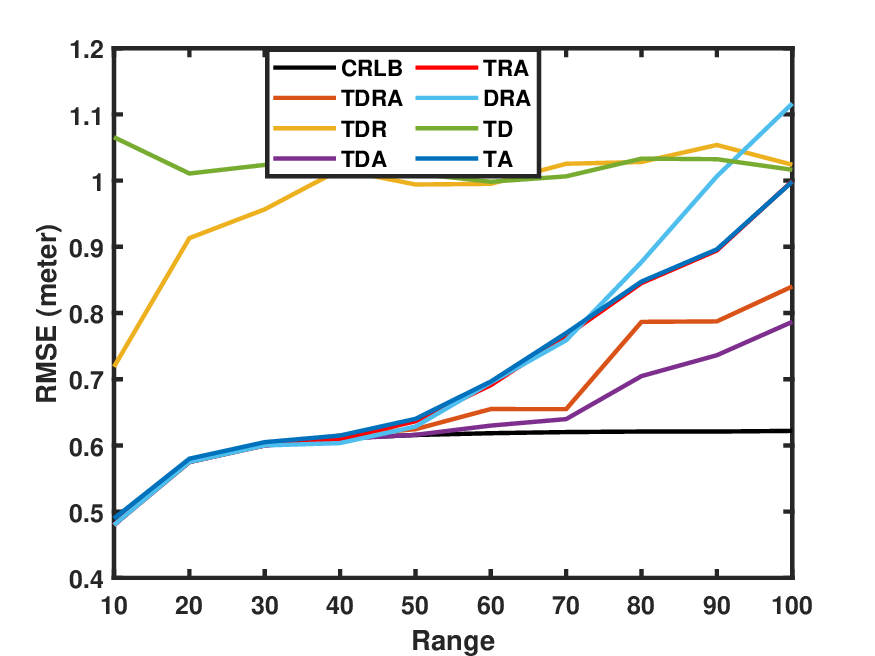}
 \caption{}
 \label{fig_rang3}
\end{subfigure}
\caption{Impact of deployment range of anchors on localization accuracy.}
\label{range}
\end{figure*}
 \begin{figure*}[ht!]
\centering
\begin{subfigure}[t]{0.47\textwidth}\centering
 \includegraphics[width=\textwidth,height=6cm]{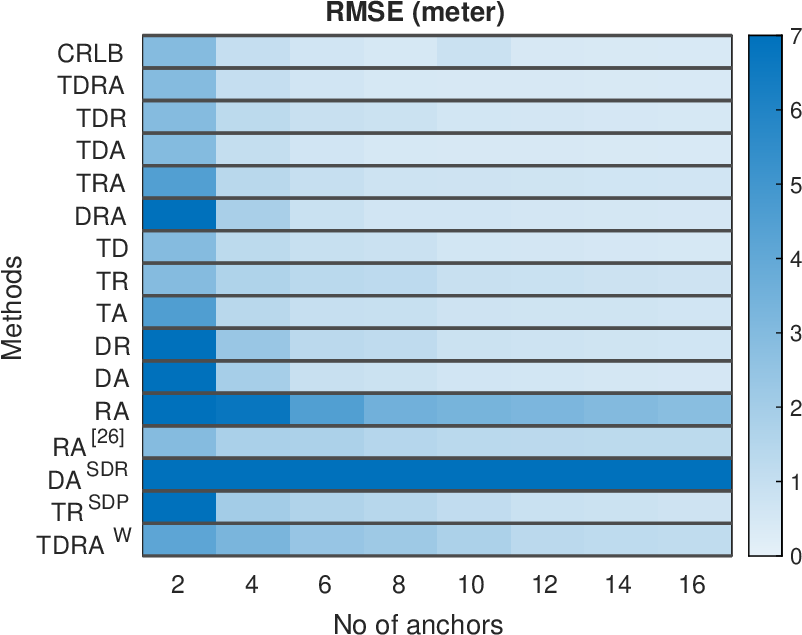}
 \label{a2}
 \caption{}
\end{subfigure}
\begin{subfigure}[t]{0.47\textwidth}\centering
 \includegraphics[width=\textwidth,height=6cm]{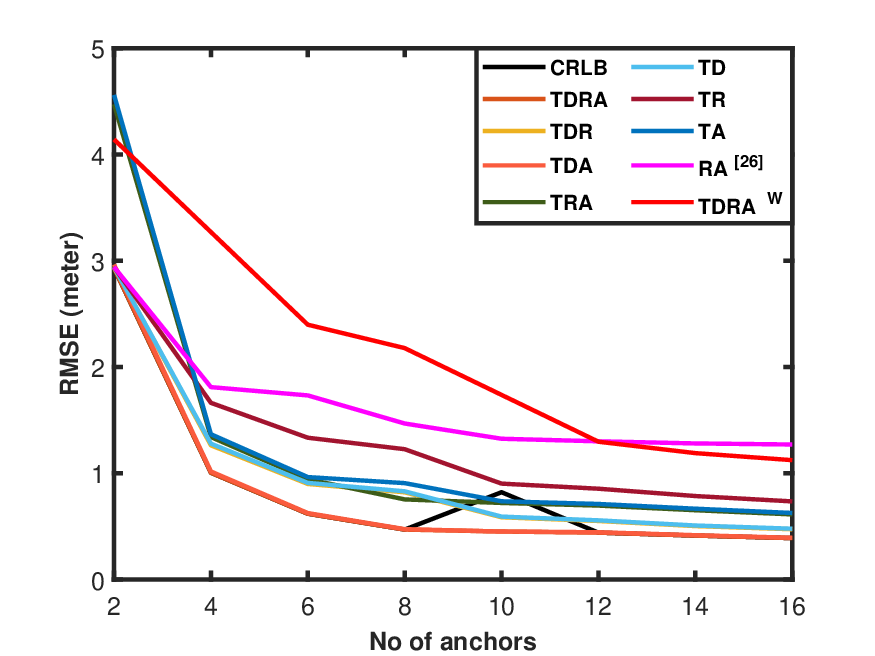}
 \label{a3}
 \caption{}
\end{subfigure}
\caption{Impact of the number of anchors on localization accuracy.}
\label{sensors}
\end{figure*} \begin{figure*}[ht!]
\centering
\begin{subfigure}[t]{0.47\textwidth}\centering
 \includegraphics[width=\textwidth,height=6cm]{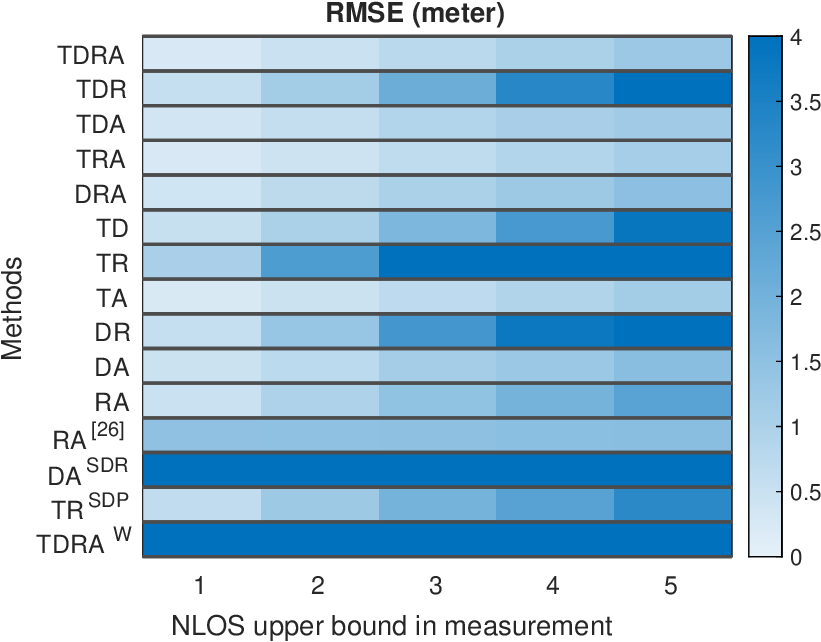}
\caption{Impact of NLOS noise in measurements.}
 \label{nl2}
\end{subfigure}
\begin{subfigure}[t]{0.47\textwidth}\centering
 \includegraphics[width=\textwidth,height=6cm]{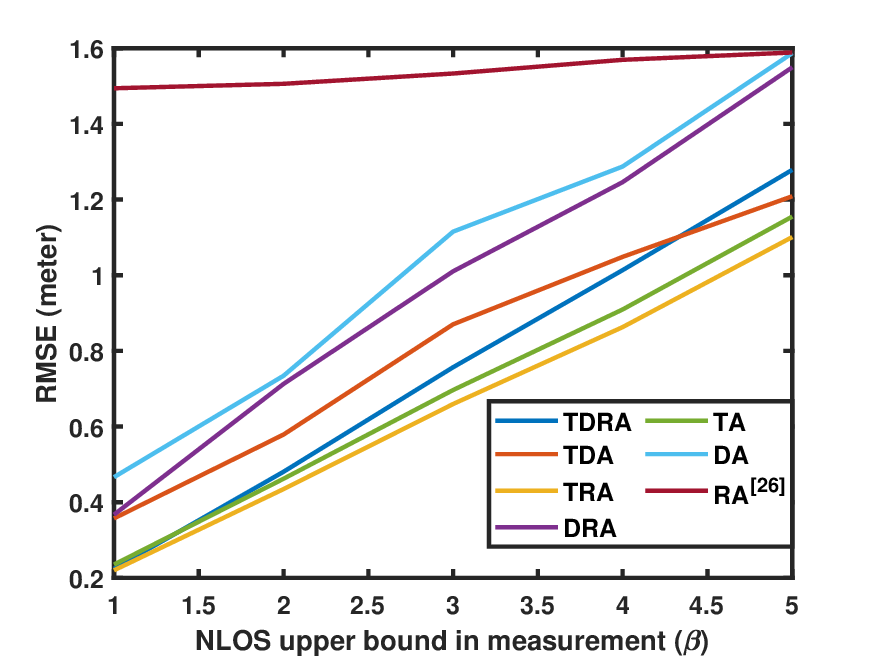}
 \caption{Impact of NLOS noise in measurements.}
 \label{nl3}
\end{subfigure}
\begin{subfigure}[t]{0.47\textwidth}\centering
 \includegraphics[width=\textwidth,height=6cm]{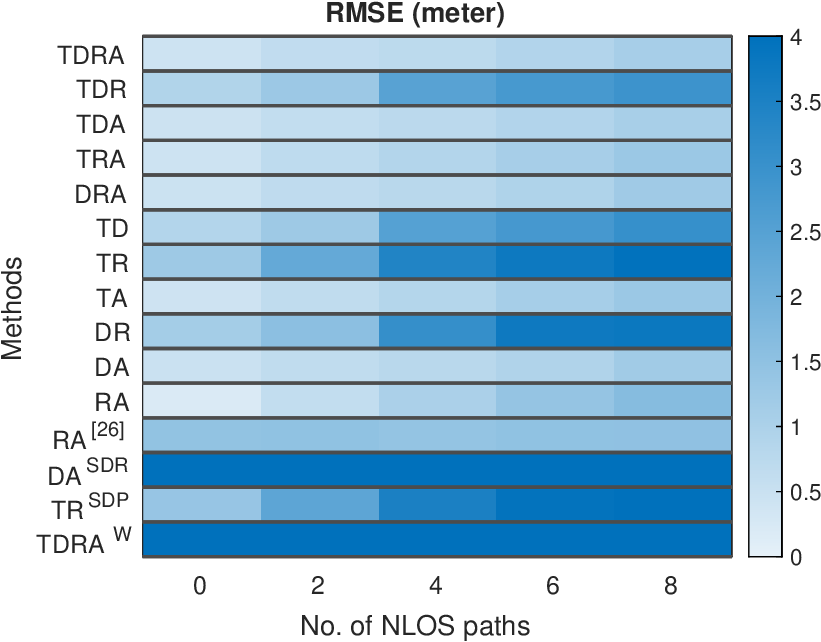}
 \caption{Impact of the number of NLOS paths in measurements.}
 \label{pl2}
\end{subfigure}
\begin{subfigure}[t]{0.47\textwidth}\centering
 \includegraphics[width=\textwidth,height=6cm]{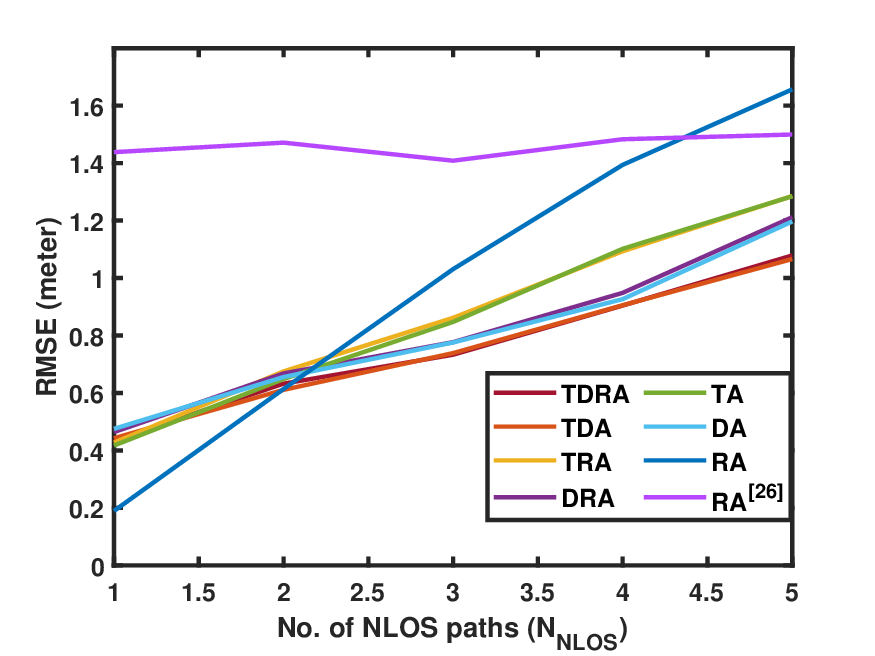}
 \caption{Impact of the number of NLOS paths in measurements.}
 \label{pl3}
\end{subfigure}
\caption{}
\label{nlos}
\end{figure*}
Since the proposed algorithm is based on the MM principle, the series of iterates obtained via the proposed method is expected to  monotonically decrease the objective function in \eqref{wls}. For the convergence analysis simulation, the configuration of the network used has the following choice of parameters  \(N=5,\hspace{2mm}R=50 \text{ meters},\hspace{2mm}\sigma=\sigma_{\scriptscriptstyle\text{TOA}}=\sigma_{\scriptscriptstyle\text{TDOA}}=\sigma_{\scriptscriptstyle\text{RSS}}=\sigma_{\scriptscriptstyle\text{AOA}}=1\). Figure \ref{fig_iter1} shows the variation of the objective of the proposed algorithm (TDRA-MM) with the iterations. Along with this, we have also shown the convergence of the proposed method based on three (TRA-MM), two (TA-MM) and one (A-MM) measurements  in Figures \ref{fig_iter2}, \ref{fig_iter3} and \ref{fig_iter4}, respectively.  As expected, the proposed algorithm and all its variants are monotonic.

% The average run-time (s) of the proposed method and the conventionally used WLS approach for the different number of anchors are listed in  table \ref{tabcc}.  It can be seen from the table that the proposed method is slower compared to the WLS approach.

\subsection{Impact of Measurement Noise}
The proposed algorithm is analyzed for variations in different parameters of the experimental setup. First, the impact of measurement noise on the localization accuracy of the proposed method is studied. In this evaluation, we have kept the  number of anchors as $N = 8$ and their deployment range as a spherical region of radius $R=50$ meters. The remaining parameters are kept the same as in the last subsection. Here, we have varied the noise in all the measurements and calculated the RMSE values. The CRLB of the hybrid TOA-TDOA-RSS-AOA is also included in the performance plots. The proposed method and its various variants (based on the type of measurements used) are compared with the WLS and other state-of-the-art methods.

First, we compare the performance of the proposed hybrid measurements with state-of-the-art methods. A heatmap of the RMSE values for all methods is presented in Figure \ref{fig_all21} at different noise variance levels. The intensity of color indicates the level of RMSE, with darker shades representing higher RMSE values and lighter shades corresponding to lower RMSE. As the variance of the noise increases, the RMSE increases. From the figure, it is evident that the RMSE values for TR (the proposed approach) are equivalent to those of the TR-SDP method. The DA-SDR method is comparable to the proposed method only at high noise levels but is less favorable when noise is low. However, these semi-definite programming-based methods exhibit high inefficiency in terms of computational complexity (with a complexity of \(\mathcal{O}(N^{4.5})\)).

The WLS method significantly lags in terms of estimation accuracy compared to the proposed algorithm. Combinations of three or more measurements (i.e., TDR, TDA, TRA, DRA, and TDRA) are generally found to be more effective than combinations of two. Next, the variation of the RMSE for the best methods out of these 15 (TDRA, TDA, TRA, DRA, TA, DA, RA, RA-[26]) with noise variance is further depicted in the next figure for a detailed study (Figure \ref{fig_all31}).

It can be observed that the RMSE values of TDA and TDRA closely match those of the CRLB (TDRA model), and the differences in RMSE values of DRA, TRA, RA and TA with CRLB are also minimal. Therefore, these combinations can be effectively used in various applications. Thus, it can be inferred that the proposed method is more accurate and computationally efficient than state-of-the-art methods. 
   \subsection{Impact of Communication Range}
In this subsection, we study the effect of variation of deployment range on the localization accuracy of the proposed algorithm and its variants. For this simulation, we have considered the number of anchors to be $N=6$. The noise in each measurement is taken to be Gaussian noise with zero mean and standard deviation $\sigma=\sigma_{\scriptscriptstyle\text{TOA}}=\sigma_{\scriptscriptstyle\text{TDOA}}=\sigma_{\scriptscriptstyle\text{RSS}}=\sigma_{\scriptscriptstyle\text{AOA}}=1$.   All the other parameters are kept same as in the previous subsection. The RMSEs of the proposed algorithm and its variants are calculated along with that of the WLS and other state-of-the-art methods.

The numerical simulations were conducted in a manner consistent with that for noise variance assessment. The results of the study on the impact of range variation are presented in Figure \ref{range}, following a similar format as discussed in Section IV C. The heatmap reveals that the RMSE values of all models increase with the expansion of the range. Notably, at any given range, the proposed method outperforms state-of-the-art methods. In the context of two-measurement hybrid scenarios, TA-MM exhibits superior performance compared to other approaches. For cases involving more than two measurements, TDRA and TDA stand out as the most effective. Furthermore, the fusion of three and four measurements demonstrates enhanced accuracy for long-range communication when compared to the fusion of only two types of measurements. For a more detailed assessment, methods with lower RMSE (TDRA, TDR, TDA, TRA, DRA, TD, TA) are selected, and their variations are depicted in Figure 5b, where all these methods closely align with the CRLB.
\begin{table*}[!t]
    \centering
      \caption{Ranking of hybrid methods for different network scenarios}
      \scalebox{0.8}{
        \begin{tabular}{|c|c|ccccccccc|}
     \hline
     \textbf{Network simulation scenario}&\textbf{Value}& \multicolumn{9}{|c|}{\textbf{Accuracy-order of the methods}}\\

    \hline
    \multirow{2}{*}{\textbf{Noise in the measurements}}&Low &TDRA&$\geq$ &TDA&$>$&DA&$>$&DRA&$>$&TRA\\
    
     &High&TDA&$\approx$ &TDRA&$>$&DA&$>$&DRA&$>$&TA \\
    \hline
    \multirow{2}{*}{\textbf{Deployment range of the sensors}}&Low &TDRA&$\geq$ &TDA&$\geq$&DA&$\geq$&RA&$>$&TRA  \\
    
     &High&TDA&$\approx$ &TDRA&$>$&TRA&$\geq$&TA&$>$&TDR \\
    \hline
    \multirow{2}{*}{\textbf{Number of sensors deployed}}&Low &TDRA&$\geq$ &TDA&$\geq$&TRA&$\geq$&TA&$>$&DA  \\
    
     &High&TDA&$\approx$ &TDRA&$>$&DA&$\geq$&DRA&$>$&TRA  \\
    \hline
    \multirow{2}{*}{\textbf{NLOS error in the measurements}}&Low &TDRA&$\geq$ &TDA&$\geq$&TRA&$\geq$&DRA&$>$&TA    \\
    
     &High&TDA&$\approx$ &TDRA&$>$&TDA&$\geq$&DRA&$>$&TA  \\
    \hline
    \end{tabular} } 
    \label{t6}
\end{table*}
\subsection{Impact of Anchor Density}
In this subsection, we study the effect of the number of anchors on the localization accuracy of the proposed algorithm and its variants. For this simulation, we have considered the deployment range of the anchors to be $R=50$ meters and the standard deviation of noise in the measurements as $\sigma=\sigma_{\scriptscriptstyle\text{TOA}}=\sigma_{\scriptscriptstyle\text{TDOA}}=\sigma_{\scriptscriptstyle\text{RSS}}=\sigma_{\scriptscriptstyle\text{AOA}}=1$. The remaining parameters are same as that of the previous simulations.  The number of anchors are varied and the RMSEs of the proposed algorithm and its variants are calculated along with that of the state-of-the-art methods.

The investigation extends to the number of anchors, and the numerical simulations are executed with a methodology akin to the previous assessments. Figure \ref{sensors} shows the simulation results, portraying the variation in RMSE concerning the number of anchors. The heatmap vividly illustrates that, as the number of anchors increases, the RMSE values across all models also witness a notable decline which is in alignment with the fact that localization efficiency is improved with the increement in number of sensors. Remarkably, the proposed method consistently outperforms state-of-the-art counterparts at varying anchor counts. Within the domain of two-measurement hybrid scenarios, TD, TR and RA-[26] maintains its superiority over other approaches. In cases involving more than two measurements, TDRA, TDR, TDA stands out as the most proficient method. The fusion of three and four measurements continues to exhibit heightened accuracy, showcasing its efficacy even when the number of anchors is too low. For a more detailed examination, methods displaying lower RMSE (TDRA, TDR, TDA, TRA, TD, TA, TR, RA-[26], TDRA$^{\text{W}}$) are singled out, and their variations are detailed in Figure 6b.  The TDRA, TRA, TDA, TDR, TD and TA methods are close to the CRLB.

\subsection{Impact of NLOS paths}
In this subsection, we study the impact of NLOS errors in the measurements on the performance of the proposed method. The NLOS noise is modeled as additional positive noise in the respective measurements listed in \cref{rss1,toa,tdoa,aoa1,aoa2}, and the probability density function of the NLOS noise is taken to be  a  uniform distribution $\mathcal{U}(0,\beta)$. For this analysis, we first consider the variation of RMSE with the NLOS noise bias ($\beta$) in the measurements shown in \cref{nl2,nl3} and next, the variation of the number of NLOS paths ($N_n$) in the simulations as shown in \cref{pl2,pl3}.  The standard deviation of noise in measurements is kept low at $\sigma=\sigma_{\scriptscriptstyle\text{TOA}}=\sigma_{\scriptscriptstyle\text{TDOA}}=\sigma_{\scriptscriptstyle\text{RSS}}=\sigma_{\scriptscriptstyle\text{AOA}}=0.1$ to study the effect of NLOS noise. From Figure \ref{nlos}, it can be concluded that the proposed method provides better accuracy when compared to other state-of-the-art methods when the measurements are corrupted by NLOS errors. In summary, the hybridization approach enhances robustness in the face of such challenges.

Finally, the findings from Section IV C-F have been consolidated, and the top five measurement combinations for each network parameter variation are listed in Table \ref{t6}. In low SNR conditions, optimal performance is achieved with combinations like DA, TA, TDA, DRA, and TDRA. For long-distance communication, hybrid methods involving time-dependent measurements (TA, TDA, TRA, TDR, and TDRA) prove favorable, while low anchor density scenarios benefit from hybrids with angle-dependent measurements (TA, DA, TRA, TDA, and TDRA). Notably, increased hybridization, such as TDA, TDR, DRA, TRA, and TDRA, significantly enhances estimation efficiency across various network scenarios, including those with low SNR, long-distance communication, and low anchor density. Moreover, these three and four-measurement combinations impart robustness to NLOS errors. In summary, the fusion of three and four measurement types ensures high accuracy, allowing for the selection of the most suitable method based on specific requirements (in terms of accuracy and cost).

\section{Conclusion}
In this paper, we have presented an efficient  algorithm for source localization using the combination of  TOA, TDOA, RSS and AOA measurements. The design objective is derived using slight approximations in the RSS and AOA models and by incorporating weights in the MLE objective. The design problem is then solved using the Majorization Minimization framework, which yields an iterative algorithm with a convergence guarantee. From the various numerical simulations conducted, the fusion of more than two measurements is found to deliver good estimation accuracy for source localization. The proposed algorithm is also quite accurate in the presence of NLOS errors in the measurements. The architecture of the proposed algorithm provides the liberty to include/exclude any type and number of measurements for localization as per the application requirements.

\section*{Declarations}
\subsection*{Ethical Approval}
Not applicable.
\subsection*{Funding}
No funding.

\subsection*{Declaration of competing interest}
The authors declare that they have no known competing financial interests or personal relationships that could have appeared to influence the work reported in this paper.
\subsection*{Availability of data and materials} Data underlying the results presented in this paper are not publicly available at this time but may be obtained from the authors upon reasonable request.

\bibliography{sn-bibliography}% common bib file

\end{document}